\documentclass{article}%
\usepackage[top=3cm, bottom=3cm, left=3cm, right=3cm]{geometry}
\usepackage{amsmath}
\usepackage{amsfonts}
\usepackage{amssymb}
\usepackage{graphicx}%
\begin{document}

\title{Non-exponential decay in quantum field theory and in quantum mechanics: the
case of two (or more) decay channels}
\author{Francesco Giacosa\\\emph{Institut f\"{u}r Theoretische Physik, Johann Wolfgang Goethe -
Universit\"{a}t}\\\emph{Max von Laue--Str. 1, D-60438 Frankfurt am Main, Germany}}
\maketitle

\begin{abstract}
We study the deviations from the exponential decay law, both in quantum field
theory (QFT) and quantum mechanics (QM), for an unstable particle which can
decay in (at least) two decay channels. After a review of general properties
of non-exponential decay in QFT and QM, we evaluate in both cases the decay
probability that the unstable particle decays in a given channel in the time
interval between $t$ and $t+dt.$ An important quantity is the ratio of the
probability of decay into the first and the second channel: this ratio is
constant in the Breit-Wigner limit (in which the decay law is exponential) and
equals the quantity $\Gamma_{1}/\Gamma_{2}$, where $\Gamma_{1}$ and
$\Gamma_{2}$ are the respective tree-level decay widths. However, in the full
treatment (both for QFT and QM) it is an oscillating function around the mean
value $\Gamma_{1}/\Gamma_{2}$ and the deviations from this mean value can be
sizable. Technically, we study the decay properties in QFT in the context of a
superrenormalizable Lagrangian with scalar particles and in QM in the context
of Lee Hamiltonians, which deliver formally analogous expressions to the QFT case.

\end{abstract}

\section{Introduction}

The study of unstable particles is a basic subject of both nuclear and
particle physics. The fact that the so called survival probability, i.e. the
probability $p(t)$ that an unstable state $S$ formed at the time $t=0$ did not
decay at the instant $t>0,$ does not follow an exponential decay law for short
times is now theoretically
\cite{ghirardi,misra,zheng,facchiprl,facchi1,balachandran} and experimentally
\cite{itano,raizen1,raizen2} established in the framework of quantum
mechanics. Moreover, oscillations on top of the exponential function can also
occur in the short time regime and could eventually last long enough to be
detected \cite{pascosc,gsiano}. Also at very large times deviations take
place: a power law, and not an exponential law, is realized \cite{ghirardi}.

In particular, in Ref. \cite{facchiprl} the deviation from the exponential
decay law has been described in the nonrelativistic quantum mechanical context
of Lee Hamiltonians \cite{lee,chiu}: this approach allows to mimic
non-relativistic quantum field theory by including a continuum of states. In
Refs. \cite{zeno1,zenoproc} these deviations have been demonstrated in the
framework of a genuinely (superrenormalizable) relativistic quantum field
theoretical Lagrangian. Both approaches give rise to qualitative similar
deviations from an exponential decay law in the short-time regime and also
show an indiction of oscillations afterwards.

In this work we study the decay of an unstable particle when more than one
decay channel is present, both in the context of relativistic quantum field
theory (QFT) and quantum mechanics (QM). This parallel allows also to show the
formal similarities of these two approaches. The equations presented in this
work are general and do not depend on the particular employed model(s).
However, it is useful for a clear presentation to work with a well-defined
theoretical setup: in QFT we employ, in line with Refs. \cite{zeno1,zenoproc},
scalar fields, which are free from spin complications, in the framework of a
superrenormalizable relativistic Lagrangian. In the case of QM we employ Lee
Hamiltonians (LH): we generalize the work of Refs. \cite{facchiprl} in order
to describe the decay of an unstable state into two distinct channels.

When (at least) two decay channels are present, one has to determine (both in
QFT and in QM) the probability of decay in each channel. Namely, the aim is to
calculate (to our knowledge for the first time) the decay probability that the
unstable particle $S$ decays in the $i$-th channel between the time interval
$t$ and $t+dt$. Each partial decay probability deviates from the corresponding
Breit-Wigner limit. In this context an important quantity under study in the
two-channel case is the ratio between the decay probability in the first
channel over the decay probability in the second channel. While this quantity
is constant in the usual Breit-Wigner (BW) limit and is equal to the ratio
$\Gamma_{1}/\Gamma_{2}$, where $\Gamma_{1}$ and $\Gamma_{2}$ are the
tree-level decay widths in the first and the second channel respectively, this
is not true in the general case. There are time intervals in which the ratio
of probabilities is larger than $\Gamma_{1}/\Gamma_{2}$, meaning that the
decay probability into the first channel is enhanced. Vice-versa, there are
time intervals in which the ratio is smaller than $\Gamma_{1}/\Gamma_{2}$,
implying that the decay probability into the second channel is enhanced.

One of the main purposes of this paper is a general and theoretically based
presentation of the equations related to the decay probability in each decay
channel as function of time. For this reason we do not refer here to any
particular physical system, but we keep the presentation as general as
possible. Yet, in order to improve the understanding of the manuscript we
shall present in the QFT case several plots of the introduced functions for a
selected numerical example. Although the employed superrenormalizable QFT
model is clearly not the most general QFT treatment, the form an the structure
of the equations are quite general and could be easily extended to other more
complicated and realistic QFT approaches.

The paper is organized as follows: In Sec. 2 we give an overview of the QFT
formalism for the description of the decay of an unstable particle. We
introduce a (superrenormalizable) Lagrangian density and discuss the relevant
properties of the survival probability $p(t)$. In\ Sec. 3 we turn to the QM
study of decays: to this end we introduce the Lee Hamiltonian, which shows a
formally very similar behavior to the QFT case of Sec. 2. We review the most
important equations in this approach and also present a relativistic
generalization of it. In Sec. 4 we turn back to QFT and discuss in detail the
formulae for the two-channel case, with special attention to the decay
probability in each channel. For sake of clarity we concentrate on the case in
which two decay channels are available; the generalization to $n$ decay
channels is straightforward and is presented in Appendix B. In QFT the
expressions of the decay probabilities in each channel cannot be analytically
derived in the most general case; to circumvent this problem, a `conjecture'
about the form of the solution is put forward. It is then shown that the
conjectured solution fulfills all the necessary requirements. A numerical
study of these solutions is presented, in order to show differences from the
usual exponential limit. In Sec. 5 we study the same problem of the
two-channel decay in QM with the help of Lee Hamiltonians: in this case it is
possible to derive the exact expressions for the decay probabilities in a
given channel and we can test the validity of the conjecture presented in the
QFT case: a numerical study shows a very good agreement of the exact results
with the proposed solutions for both the nonrelativistic and relativistic
treatments of the Lee Hamiltonians. Finally, in Sec. 6 we briefly present our
conclusions and outlooks.

\section{General properties of decays in quantum field theory}

\subsection{A simple Lagrangian to study the two-channel problem}

We introduce a relativistic (superrenormalizable) Lagrangian with the three
scalar fields $S$, $\varphi_{1}$ and $\varphi_{2}$:
\begin{equation}
\mathcal{L}=\frac{1}{2}(\partial_{\mu}S)^{2}-\frac{1}{2}M_{0}^{2}S^{2}%
+\frac{1}{2}(\partial_{\mu}\varphi_{1})^{2}-\frac{1}{2}m_{1}^{2}\varphi
_{1}^{2}+\frac{1}{2}(\partial_{\mu}\varphi_{2})^{2}-\frac{1}{2}m_{2}%
^{2}\varphi_{2}^{2}+g_{1}S\varphi_{1}^{2}+g_{2}S\varphi_{2}^{2}\text{ .}
\label{lag}%
\end{equation}
For definiteness, we assume that $m_{1}<m_{2}$. The interaction terms
$g_{1}S\varphi_{1}^{2}$ and $g_{2}S\varphi_{2}^{2}$ induce the decay processes
$S\rightarrow\varphi_{1}\varphi_{1}$ and $S\rightarrow\varphi_{2}\varphi_{2}$,
see the pictorial representation in Fig. 1, panel (a).

It is useful to introduce the tree-level decay function as%
\begin{equation}
\Gamma^{\text{t-l}}(x,m,g)=\frac{\sqrt{\frac{x^{2}}{4}-m^{2}}}{8\pi x^{2}%
}(\sqrt{2}g)^{2}\theta(x-2m)\text{ ,} \label{tldf}%
\end{equation}
where $\theta(x)$ is the step function and $x$ plays the role of the `running'
mass of the unstable state $S$. The tree-level decay functions in the first
and the second channel read respectively%
\begin{align}
\Gamma_{S\varphi_{1}\varphi_{1}}^{\text{t-l}}(x)  &  =\Gamma^{\text{t-l}%
}(x,m_{1},g_{1})=\frac{\sqrt{\frac{x^{2}}{4}-m_{1}^{2}}}{8\pi x^{2}}(\sqrt
{2}g_{1})^{2}\theta(x-2m_{1})\text{ ,}\nonumber\\
\Gamma_{S\varphi_{2}\varphi_{2}}^{\text{t-l}}(x)  &  =\Gamma^{\text{t-l}%
}(x,m_{2},g_{2})=\frac{\sqrt{\frac{x^{2}}{4}-m_{2}^{2}}}{8\pi x^{2}}(\sqrt
{2}g_{2})^{2}\theta(x-2m_{2})\text{ .} \label{tl1}%
\end{align}
Similarly, we also introduce the total tree-level decay function as
\begin{equation}
\Gamma_{S}^{\text{t-l}}(x)=\Gamma_{S\varphi_{1}\varphi_{1}}^{\text{t-l}%
}(x)+\Gamma_{S\varphi_{2}\varphi_{2}}^{\text{t-l}}(x)\text{ .} \label{tlstot}%
\end{equation}

The partial tree-level decay rates --as calculated from the Lagrangian
(\ref{lag})-- are naively obtained by setting $x=M_{0}$ into Eqs. (\ref{tl1}).
However, care is needed: as shown in the next subsection, the mass of the
unstable particle $S$ is modified by quantum
fluctuations\footnote{Alternatively, one could introduce a counterterm and
work with $M_{0}$ as the physical mass, see details in Appendix 1.} (it is
typically shifted to lower values):%
\begin{equation}
M_{0}\rightarrow M\text{ .}%
\end{equation}
For this reason the on-shell tree-level expressions $\Gamma_{1}$ and
$\Gamma_{2}$ for the partial decay widths are not evaluated at $M_{0}$, but at
the physical value $M$:
\begin{equation}
\Gamma_{1}=\Gamma_{S\varphi_{1}\varphi_{1}}^{\text{t-l}}(x=M)\text{ and
}\Gamma_{2}=\Gamma_{S\varphi_{2}\varphi_{2}}^{\text{t-l}}(x=M)\text{ .}%
\end{equation}
Then, the total on-shell tree-level decay width reads
\begin{equation}
\Gamma=\Gamma_{S}^{\text{t-l}}(x=M)=\Gamma_{1}+\Gamma_{2}\text{ .}%
\end{equation}
As a consequence, the tree-level (or Breit-Wigner) expression of the survival
probability $p(t)$ for the resonance $S$ created at $t=0$ is the usual
exponential function
\begin{equation}
p_{\text{BW}}(t)=e^{-\Gamma t}\text{ ,}%
\end{equation}
and the corresponding tree-level expression of the mean lifetime is
$\tau_{\text{BW}}=1/\Gamma$. The explicit derivation of the exponential form
is obtained through the Breit-Wigner limit in Sec. 2.3.%

\begin{figure}
[ptb]
\begin{center}
\includegraphics[
height=2.5598in,
width=5.9499in
]%
{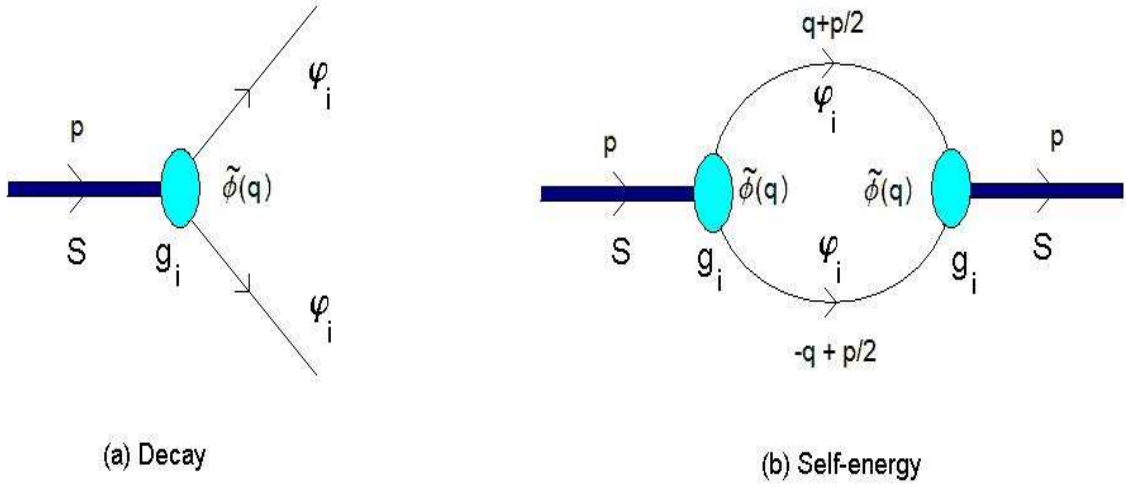}%
\caption{Pictorial representation of (a) the amplitude of the tree-level decay
process $S\rightarrow\varphi_{i}\varphi_{i}$ with $i=1,2$ and (b) the
corresponding self-energy diagram. The imaginary part of (b) is related to
(a), see Eq. (\ref{impi}) and the discussion in the text. The quantity
$\tilde{\phi}(q)$ is the vertex function, which is chosen to be $\tilde{\phi
}(q)=\theta(\Lambda^{2}-\mathbf{q}^{2})$ for the numerical evaluations.}%
\end{center}
\end{figure}

\bigskip

\subsection{The mass distribution $d_{S}(x)$ and the survival probability
$p(t)$}

A crucial intermediate step toward the determination of the survival
probability $p(t)$ in the framework of quantum field theory is the evaluation
of the propagator $G_{S}(p^{2})$ of the unstable resonance $S.$ The quantity
$G_{S}(p^{2})$ is obtained at the 1-loop level by (re)summing the one-particle
irreducible self-energy contribution (Fig. 1, panel (b)):%
\begin{equation}
G_{S}(p^{2})=\left[  p^{2}-M_{0}^{2}+(\sqrt{2}g_{1})^{2}\Sigma(p^{2},m_{1}%
^{2})+(\sqrt{2}g_{2})^{2}\Sigma(p^{2},m_{2}^{2})+i\varepsilon\text{ }\right]
^{-1}\text{,} \label{propqft}%
\end{equation}
where $\Sigma(p^{2},m^{2})$ is the amplitude of the standard loop diagram in
which two virtual particles of the $\varphi$-type circulate:
\begin{equation}
\Sigma(p^{2},m^{2})=-i\int\frac{d^{4}q}{(2\pi)^{4}}\frac{1}{\left[  \left(
\frac{p}{2}+q\right)  ^{2}-m^{2}+i\varepsilon\right]  \left[  \left(  \frac
{p}{2}-q\right)  ^{2}-m^{2}+i\varepsilon\right]  }\text{ .} \label{self}%
\end{equation}
It is also useful for future purposes to introduce the function $\Pi(p^{2})$
as%
\begin{equation}
\Pi(p^{2})=(\sqrt{2}g_{1})^{2}\Sigma(p^{2},m_{1}^{2})+(\sqrt{2}g_{2}%
)^{2}\Sigma(p^{2},m_{2}^{2})\text{ ,}%
\end{equation}
in such a way that the propagator is given by the simple expression:
\begin{equation}
G_{S}(p^{2})=\left[  p^{2}-M_{0}^{2}+\Pi(p^{2})+i\varepsilon\text{ }\right]
^{-1}\text{.}%
\end{equation}
Due to the optical theorem the imaginary part of the function $\Pi(p^{2}%
=x^{2})$ is related to the tree-level decay function in Eq. (\ref{tlstot}):
\begin{equation}
\operatorname{Im}\Pi(x^{2})=x\Gamma_{S}^{\text{t-l}}(x)\text{ .} \label{impi}%
\end{equation}
Although the imaginary part is convergent, this is not the case for the real
part. For this reason the loop function $\Sigma(p^{2},m^{2})$ must be
regularized: in this work we use a sharp, three-dimensional cutoff $\Lambda$
for the explicit numerical examples. However, different choices of the
regularization procedures do not change the qualitative features of the
discussion \cite{zeno1,zenoproc,lupo1}. The explicit expression for the
function $\Sigma(p^{2},m^{2})$ and more details on the consequences of the
optical theorem are given in Appendix A.

The spectral function $d_{S}(x)$ of the scalar field $S$ is defined as the
imaginary part of the propagator:%
\begin{equation}
d_{S}(x=\sqrt{p^{2}})=\frac{2x}{\pi}\left\vert \lim_{\varepsilon\rightarrow
0}\operatorname{Im}[G_{S}(x^{2})]\right\vert \text{ .} \label{ds}%
\end{equation}
Explicitly:%
\begin{equation}
d_{S}(x)=\frac{2x}{\pi}\lim_{\varepsilon\rightarrow0}\frac{\operatorname{Im}%
\left[  \Pi(x^{2})\right]  +\varepsilon}{\left(  x^{2}-M_{0}^{2}%
+\operatorname{Re}\Pi(x^{2})\right)  ^{2}+\left(  \operatorname{Im}\Pi
(x^{2})+\varepsilon\right)  ^{2}}\text{ .} \label{dsx}%
\end{equation}
The renormalized mass $M$ of the particle $S$ is defined as the zero of the
real part of $G_{S}(p^{2})^{-1}:$%
\begin{equation}
M^{2}-M_{0}^{2}+\operatorname{Re}\Pi(M^{2})=0\text{ .} \label{eqm}%
\end{equation}
Usually, $\operatorname{Re}\Pi(M^{2})>0,$ thus the renormalized mass $M$ is
smaller than $M_{0}.$

The quantity $d_{S}(x)dx$ represents the probability that in the rest frame of
$S$ the state $S$ has a mass between $x$ and $x+dx.$ For this reason it is
also denoted as the mass distribution of the unstable state $S$. It is
correctly normalized for each $g_{1},g_{2}$,
\begin{equation}
\int_{0}^{\infty}\mathrm{dx}d_{S}(x)=1\text{ ,} \label{norm}%
\end{equation}
and, in the limit $g_{1}\rightarrow0$ and $g_{2}\rightarrow0,$ the expected
spectral function $d_{S}(x)=\delta(x-M_{0})$ is obtained, see details in Refs.
\cite{lupo1,achasov}.

In Fig. 2 we present the spectral function $d_{S}(x)$ for a selected numerical
example, where we chose as energy unit the quantity $\left[  2m_{2}\right]
$:
\begin{equation}
M_{0}=\frac{3}{2}\left[  2m_{2}\right]  \text{ , }g_{1}=g_{2}=3\left[
2m_{2}\right]  \text{ , }m_{1}=\frac{1}{10}\left[  2m_{2}\right]  \text{ ,
}m_{2}=\frac{1}{2}\left[  2m_{2}\right]  \text{ , }\Lambda=2\left[
2m_{2}\right]  \text{ .} \label{numval}%
\end{equation}
As a result one obtains from Eq. (\ref{eqm}) that $M=1.40\left[
2m_{2}\right]  $ (which is just slightly smaller than $M_{0}$: the quantum
fluctuations are not large in this example). The tree-level decay widths read:
$\Gamma_{1}=0.13\left[  2m_{2}\right]  $, $\Gamma_{2}=0.089\left[
2m_{2}\right]  $ and $\Gamma=0.22\left[  2m_{2}\right]  .$ As a consequence,
the tree-level lifetime reads $\tau_{\text{BW}}=4.63\left[  2m_{2}\right]
^{-1}$. We shall use these numerical values in all the upcoming plots of the
paper, because they allow for a clear visualization of the underlying
equations. It should be however stressed that the qualitative behavior is not
dependent on the precise numerical choice.%

\begin{figure}
[ptb]
\begin{center}
\includegraphics[
height=2.6852in,
width=4.0871in
]%
{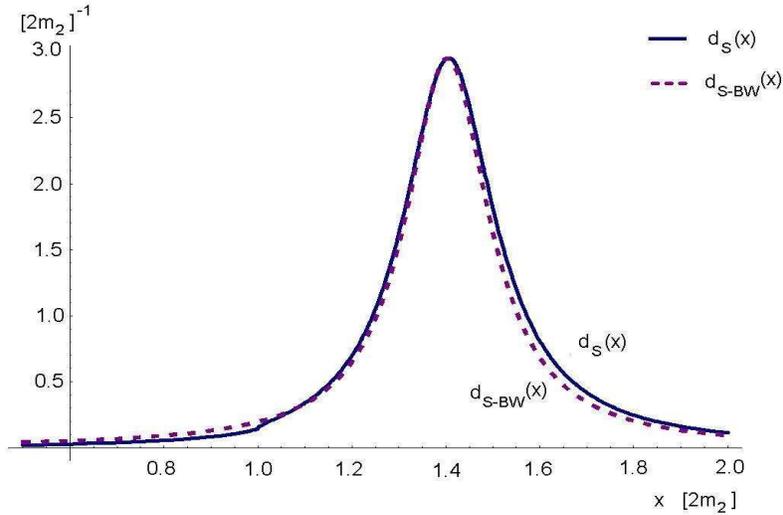}%
\caption{The spectral function $d_{S}(x)$ from Eq. (\ref{ds}) and the
nonrelativistic BW limit $d_{S\text{-BW}}(x)$, see Eq. (\ref{dbw}) derived in
Sec. 2.3, are plotted for the numerical values of Eq. (\ref{numval}). In this
example the function $d_{S}(x)$ differs only slightly from the BW limit, but
this small deviation is enough to generate sizable deviations in the early
temporal evolution, see later on.}%
\end{center}
\end{figure}

It is now possible to determine the probability amplitude $a(t)$, and
therefore the survival probability $p(t)$. To this end, let us consider the
set $\left\{  \left\vert x\right\rangle \right\}  $ of eigenstates of the full
Hamiltonian $H$ of the system defined by the Lagrangian of Eq. (\ref{lag}):
$H\left\vert x\right\rangle =x\left\vert x\right\rangle .$ The continuous
variable $x$ describes two-particle states and therefore is such that
$x\geq2m_{1}.$ The survival amplitude $a(t)$ reads (in the Heisenberg picture)%
\begin{equation}
a(t)=\left\langle S\left\vert e^{-iHt}\right\vert S\right\rangle =\left\langle
S|x\right\rangle e^{-ixt}\left\langle x|S\right\rangle =\left\vert
\left\langle S|x\right\rangle \right\vert ^{2}e^{-ixt}\text{ ,} \label{a(t)}%
\end{equation}
where the integration over the continuos variable $x$ is understood. Taking
into account that $\left\vert \left\langle S|x\right\rangle \right\vert
^{2}=d_{S}(x)$ (i.e., as shown above, $d_{S}(x)dx$ is the probability that the
unstable state $S$ has an energy between $x$ and $x+dx$) and making the
integration explicit, we get:%

\begin{equation}
a(t)=\int_{-\infty}^{\infty}\mathrm{dx}d_{S}(x)e^{-ixt}\text{ , }%
p(t)=\left\vert a(t)\right\vert ^{2}\text{ .} \label{p(t)}%
\end{equation}
The integrand in Eq. (\ref{p(t)}) is nonzero only between $(2m_{1},\infty)$ in
virtue of the step function in Eq. (\ref{impi}). However, it is from a
technical point of view helpful to write the integration range as
$(-\infty,\infty)$ so that the quantity $a(t)$ is expressed as the Fourier
transform of the spectral function $d_{S}(x).$ It is in this way that the
usual exponential law can be derived in the so-called Breit-Wigner limit, see
Sec. 2.3.

The condition $p(0)=1$ is fulfilled in virtue of the normalization of
$d_{S}(x).$ This property is, in turn, a consequence of the resummation and of
the validity of the so-called K\"{a}llen-Lehman representation \cite{achasov}.
In Fig. 3 the function $p(t)$ and the corresponding exponential limit
$p_{\text{BW}}(t)=e^{-\Gamma t}$ are plotted: it is evident that sizable
deviations from the `full' result $p(t)$ take place\footnote{Here `full'
result means that the resummed 1-loop approximation has been performed and
that the full functional form of the spectral function $d_{S}(x)$ is kept in
the evaluation of the temporal behavior.}. This fact is remarkable if one
considers that the deviations of the spectral function $d_{S}(x)$ from the
Breit-Wigner limit $d_{S\text{-BW}}(x)$ are very small, see Fig. 2.
Nevertheless, these small deviations are enough to generate large changes in
the temporal evolution of the system: for a better visualization of this
point, the difference $p(t)-p_{\text{BW}}(t)$ has been also plotted in Fig. 3.

The spectral function $d_{S}(x)$ is naturally bound from below because of the
presence of the low-energy threshold $2m_{1}.$ Its behavior for large energies
$x$ can be easily inferred by using the optical theorem in\ Eq. (\ref{impi}):
$d_{S}(x)\sim x^{-3}$ for large $x$. Then, the presence of cutoff $\Lambda$
implies that $d_{S}(x)=0$ for $x>2\sqrt{\Lambda^{2}+m_{1}^{2}}\simeq2\Lambda.$
It thus follows that the mean energy
\begin{equation}
\left\langle x\right\rangle =\int_{-\infty}^{\infty}\mathrm{dx}xd_{S}(x)
\end{equation}
is a finite number. As shown in Ref. \cite{ghirardi} the existence of an
energy threshold and the finiteness of $\left\langle x\right\rangle $ suffice
to guarantee short-time (and also and long-time) deviations from the
exponential decay law. Quite remarkably, in the present superrenormalizable
model of Eq. (\ref{lag}) the finiteness of $\left\langle x\right\rangle $ is
guaranteed also in the limit $\Lambda\rightarrow\infty,$ being $xd_{S}(x)\sim
x^{-2}$ for large $x.$ On the contrary, all higher momenta $\left\langle
x^{n}\right\rangle =\int_{-\infty}^{\infty}\mathrm{dx}x^{n}d_{S}(x)$ with
$n=2,3,...$diverge for $\Lambda\rightarrow\infty$. As shown in Ref.
\cite{napa}, the finiteness of $\left\langle x\right\rangle $ also assures the
differentiability of the amplitude $a(t)$ (and thus also of $p(t)$), which we
will use in the following considerations and also in Secs. 2.4 and 2.5 when
discussing in more detail the short-time behavior and the quantum-Zeno effect.%

\begin{figure}
[ptb]
\begin{center}
\includegraphics[
height=2.8911in,
width=4.5602in
]%
{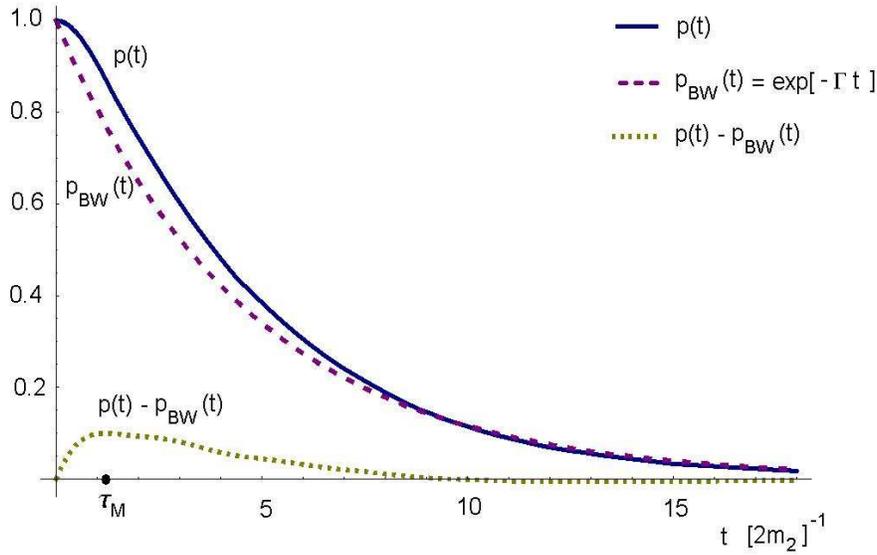}%
\caption{The survival probability $p(t)$, the BW limit $p_{\text{BW}%
}(t)=e^{-\Gamma t}$ and the difference $p(t)-p_{\text{BW}}(t)$ are plotted for
the numerical values of Eq. (\ref{numval}). The dot on the horitonzal axis
corresponds to the value of $\tau_{M}$, see Eq. (\ref{taum}) and the
discussion in Sec. 2.4.}%
\end{center}
\end{figure}

We now introduce the decay probability density $h(t)$: the quantity $h(t)dt$
represents the probability that the unstable particle $S$ decays in the time
interval $(t,t+dt)$. Clearly, the integrated quantity $w(t)$ defined as
\begin{equation}
w(t)=\int_{0}^{t}\mathrm{du}h(u)=1-p(t)
\end{equation}
represents the probability that the unstable particle decays between $0$ and
$t.$ Deriving the latter equation we obtain%
\begin{equation}
h(t)=-\frac{dp}{dt}=-p^{\prime}(t)\text{ ,} \label{h}%
\end{equation}
i.e. $h(t)$ is simply the negative derivative of the survival probability
$p(t)$.

In terms of state vectors the probability $w(t)$ can be symbolically expressed
as%
\begin{equation}
w(t)=\int_{0}^{t}\mathrm{du}h(u)=\sum_{i=1,2}\sum_{\mathbf{k}}\left\vert
\left\langle \varphi_{i,\mathbf{k}}\varphi_{i,-\mathbf{k}}\left\vert
e^{-iHt}\right\vert S\right\rangle \right\vert ^{2}\text{ ,} \label{intothsv}%
\end{equation}
where the state $\left\vert \varphi_{i,\mathbf{k}}\varphi_{i,-\mathbf{k}%
}\right\rangle $ represents the state of two particles of the type
$\varphi_{i}$, which carry opposite three-momentum $\mathbf{k}$ and
$-\mathbf{k,}$ respectively. In fact, the object $\left\vert \left\langle
\varphi_{i,\mathbf{k}}\varphi_{i,-\mathbf{k}}\left\vert e^{-iHt}\right\vert
S\right\rangle \right\vert ^{2}$ is the probability that the particle $S$
decays into two particles of the type $\varphi_{i}$, the first with momentum
$\mathbf{k}$ and the second with momentum $-\mathbf{k.}$ The overall decay
probability $w(t)$ results as the sum of all these (independent)
probabilities. It is important to stress that in Eq. (\ref{intothsv}) we
neglect the decays into $n$-particle states with $n=4,6,8,..,$ which are
suppressed by kinematical factors and anyhow neglected in the self-energy
contributions. Note also that the summation $\sum_{\mathbf{k}}$ means the sum
over all the momenta $\mathbf{k}=2\pi\mathbf{n}/L,$ where $\mathbf{n}%
=(n_{1},n_{2},n_{3})$ and $n_{i}=0,\pm1,\pm2,...$; the system is thus confined
into a box with volume $V=L^{3}$. The limit $L\rightarrow\infty$ is
straightforward but unnecessary for the present discussion (see Sec. 3.1 for
the explicit evaluation of it).

Upon deriving Eq. (\ref{intothsv}) we can also express $h(t)$ in terms of
state vectors:%
\begin{equation}
h(t)=\frac{d}{dt}\left(  \sum_{i=1,2}\sum_{\mathbf{k}}\left\vert \left\langle
\varphi_{i,\mathbf{k}}\varphi_{i,-\mathbf{k}}\left\vert e^{-iHt}\right\vert
S\right\rangle \right\vert ^{2}\right)  \text{ .} \label{hsv}%
\end{equation}
It is indeed a remarkable fact that, in order to evaluate $h(t)$ we do not
need to evaluate explicitly the matrix elements $\left\langle \varphi
_{i,\mathbf{k}}\varphi_{i,-\mathbf{k}}\left\vert e^{-iHt}\right\vert
S\right\rangle $, but we simply have to perform the derivative of Eq. (\ref{h}).

In Fig. 4 the density of decay probability $h(t)$ and its corresponding BW
limit $h_{\text{BW}}(t)=-p_{\text{BW}}^{\prime}(t)=\Gamma e^{-\Gamma t}$ are
plotted. Here the difference between the `full' result and the BW limit are
even more evident, due to the fact that $h(t\rightarrow0)$ vanishes and
$h_{\text{BW}}(t\rightarrow0)=\Gamma$ clearly does not, see Sec. 2.4 for a
detailed description of this point. Moreover, one can clearly see the presence
of oscillations of the function $h(t)$, in agreement with the discussion of
Ref. \cite{gsiano}.%

\begin{figure}
[ptb]
\begin{center}
\includegraphics[
height=2.7389in,
width=4.4823in
]%
{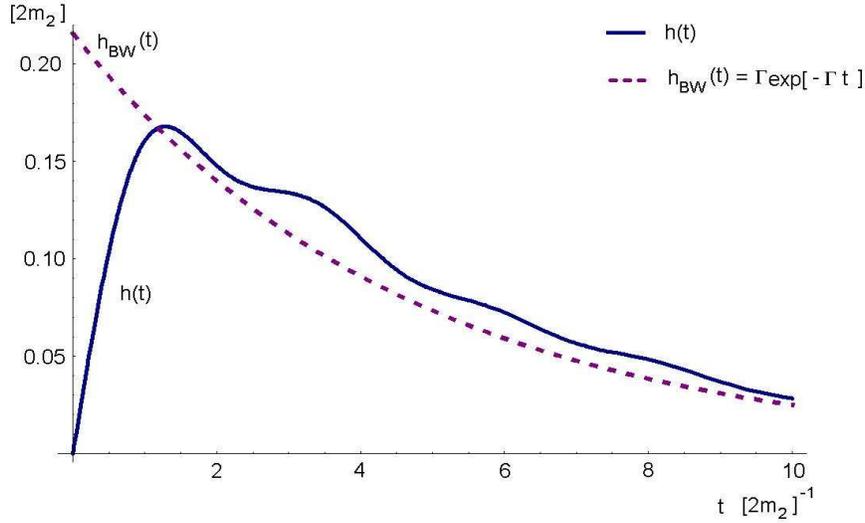}%
\caption{ The decay probability density $h(t)$ and the corresponding BW limit
$h_{\text{BW}}(t)=\Gamma e^{-\Gamma t}$ are plotted for the numerical choice
of Eq. (\ref{numval}). For short times the two functions are qualitatively
(and here also quantitatively) very different, while the former vanishes and
the latter does not. Moreover, oscillations on top of the exponential are
visible. Then, for large times the function $h(t)$ approaches the BW limit
$h_{\text{BW}}(t)$. }%
\end{center}
\end{figure}

\subsection{The Breit-Wigner limit}

It is instructive to discuss the Breit-Wigner limit, in which the survival
probability $p(t)$ turns out to be exponential. We start by approximating
$d_{S}(x)$ as follows:
\begin{equation}
d_{S}(x)=\frac{2x}{\pi}\frac{x\Gamma_{S}^{\text{t-l}}(x)}{\left(  x^{2}%
-M_{0}^{2}+\operatorname{Re}\Pi(x^{2})\right)  ^{2}+\left(  x\Gamma
_{S}^{\text{t-l}}(x)\right)  ^{2}}\simeq\frac{2M}{\pi}\frac{M\Gamma
_{S}^{\text{t-l}}(M)\theta(x-2m_{1})}{\left(  x^{2}-M^{2}\right)
^{2}+(M\Gamma_{S}^{\text{t-l}}(M))^{2}}\text{ .} \label{bwrelapp}%
\end{equation}
By having included the step function $\theta(x-2m_{1})$ we have kept track of
the threshold. A subtle point of Eq. (\ref{bwrelapp}) is the normalization,
which is spoiled by the employed approximation. This property is however
crucial in order to have $p(0)=1.$ It is thus necessary to rescale the
function in such a way that it is normalized to $1$. Finally, the spectral
function in the relativistic BW approximation reads:%
\begin{equation}
d_{S\text{-BWrel}}(x)=\frac{N_{\text{BWrel}}}{\left(  x^{2}-M^{2}\right)
^{2}+(M\Gamma)^{2}}\theta(x-2m_{1})\text{ ,}%
\end{equation}
where%
\begin{equation}
N_{\text{BWrel}}=\left[  \int_{-\infty}^{\infty}\mathrm{dx}\frac
{\theta(x-2m_{1})}{\left(  x^{2}-M^{2}\right)  ^{2}+(M\Gamma)^{2}}\right]
^{-1}=\left[  \int_{2m_{1}}^{\infty}\mathrm{dx}\frac{1}{\left(  x^{2}%
-M^{2}\right)  ^{2}+(M\Gamma)^{2}}\right]  ^{-1}\text{ .}%
\end{equation}
A further simplification, leading to the usual, nonrelativistic Breit-Wigner
form, is needed:%
\[
d_{S\text{-BWrel}}(x)=\frac{N_{\text{BWrel}}\theta(x-2m_{1})}{\left(
x^{2}-M^{2}\right)  ^{2}+(M\Gamma)^{2}}=\frac{N_{\text{BWrel}}\theta
(x-2m_{1})}{\left(  x-M\right)  ^{2}(x+M)^{2}+(M\Gamma)^{2}}%
\]%
\begin{equation}
\simeq\frac{N_{\text{BWrel}}\theta(x-2m_{1})}{\left(  x-M\right)  ^{2}%
(2M)^{2}+(M\Gamma)^{2}}\simeq\frac{N_{\text{BWrel}}}{\left(  x-M\right)
^{2}(2M)^{2}+(M\Gamma)^{2}}\text{.}%
\end{equation}
Note that in the last equation we have also dropped the requirement that
$x>2m_{1}$, thus $x$ can now vary from $-\infty$ to $+\infty.$ This is of
course unphysical for the present Hamiltonian because negative values of $x$,
i.e. negative values of the running mass of the unstable state $S$, do not
make sense. Nevertheless, the function is peaked at $M>0$ and the extension to
negative $x$ affects the function for $x>2m_{1}$ only negligibly\footnote{This
is true for the nonrelativistic approximation, but is not true for the
relativistic one: in fact, the latter is peaked also for $x=-M.$}. As a
result, the spectral function in the nonrelativistic BW approximation reads%
\begin{equation}
\text{ }d_{S\text{-BW}}(x)=\frac{N_{\text{BW}}}{\left(  x-M\right)
^{2}+\Gamma^{2}/4}\text{ ,} \label{dbw}%
\end{equation}
where $N_{\text{BW}}=\Gamma/2\pi$ is the normalization constant such that
$\int_{-\infty}^{\infty}\mathrm{dx}d_{S\text{-BW}}(x)=1.$

The amplitude $a(t)$ as calculated from Eqs. (\ref{p(t)}) and (\ref{dbw})
takes the form%
\begin{equation}
a_{\text{BW}}(t)=e^{-\Gamma t/2}e^{-iMt}\text{ (}t>0\text{)} \label{abw}%
\end{equation}
which implies the exponential decay law%
\begin{equation}
p_{\text{BW}}(t)=\left\vert a_{\text{BW}}(t)\right\vert ^{2}=e^{-\Gamma
t}\text{ .} \label{pbw}%
\end{equation}
Notice that this result has been obtained by extending the range of
integration in Eq. (\ref{p(t)}) to $(-\infty,\infty)$. Obviously, the
Breit-Wigner distribution is never exact and cannot be fulfilled by any
physical system. It is in this sense remarkable that deviations from the
exponential decay law have been discussed only relatively recently. This has
surely to do with the fact that the BW distribution, although never exact,
works extremely well for the majority of physical systems in a sufficiently
large energy interval.\ Conversely, as function of time the deviations from
the exponential law take place very early and very late, and are thus
difficult to measure.

The decay probability density $h(t)$ defined in Eq. (\ref{h}) shows also an
exponential behavior in the BW limit:%
\begin{equation}
h(t)\rightarrow h_{\text{BW}}(t)=-p_{\text{BW}}^{\prime}(t)=\Gamma e^{-\Gamma
t}\text{ .}%
\end{equation}
The exponential functions $p_{\text{BW}}(t)$ and $h_{\text{BW}}(t)$ have been
plotted and compared with the `full' results $p(t)$ and $h(t)$ in Fig. 3 and
4, respectively.

\subsection{Vanishing of $p^{\prime}(0)$ and the definition of the
`Zeno-time'}

Let us consider the first derivative of the survival probability $p(t)$ for
$t=$ $0,$ $p^{\prime}(0)$. To this end we study the quantity%
\begin{equation}
a^{\prime}(t)=\int_{-\infty}^{\infty}\mathrm{dx}(-ix)d_{S}(x)e^{-ixt}\text{ .}%
\end{equation}
In virtue of the cutoff (independently on its value) the integral
$\int_{-\infty}^{\infty}\mathrm{dx}xd_{S}(x)e^{-ixt}$ is well defined and does
not contain any divergences. In particular
\begin{equation}
a^{\prime}(t=0)=-i\int_{-\infty}^{\infty}\mathrm{dx}xd_{S}(x)=-i\left\langle
x\right\rangle
\end{equation}
is a well defined imaginary number, being $\left\langle x\right\rangle $ a
finite real number, as discussed in Sec. 2.2.

The function $p^{\prime}(t)$ can be expressed as%
\begin{equation}
p^{\prime}(t)=a^{\prime\ast}(t)a(t)+a^{\ast}(t)a^{\prime}(t)\text{ .}%
\end{equation}
Being $a(0)=1$ we obtain%
\begin{equation}
p^{\prime}(0)=i\left\langle x\right\rangle -i\left\langle x\right\rangle =0.
\end{equation}
The first derivative of the survival probability vanishes for $t=0.$ This
means that for small times a deviation from the exponential law, for which one
obviously has $\lim_{t\rightarrow0^{+}}p_{BW}^{\prime}(t)=-\Gamma$ (see Eq.
(\ref{pbw})), is realized.

The natural question is how long does the deviation form the exponential law
lasts. In the literature a Taylor expansion has been often performed,%
\begin{equation}
p(t)=1-\frac{t^{2}}{\tau_{Z}^{2}}+...\text{ ,} \label{tauzenoold}%
\end{equation}
where the coefficient
\begin{equation}
\tau_{Z}=\sqrt{\frac{1}{\left\langle x^{2}\right\rangle -\left\langle
x\right\rangle ^{2}}} \label{tauzeno}%
\end{equation}
has been defined as the `Zeno time', which quantifies the time interval of
non-exponential behavior. However, as already explained in Ref. \cite{zeno1},
the second order derivative does not even need to exist. As a trivial
illustrative example one can consider the simple and well defined function
$p(t)=1-t^{3/2}$, for which $p^{\prime}(0)=0$ but $p^{(n>1)}(0)=\infty.$
Moreover, even when the second derivative exists --as in all physical cases,
see below-- this does not necessarily mean that the parameter $\tau_{Z}$ in
Eq. (\ref{tauzenoold}) represents a correct estimate of the time interval in
which a deviation from the exponential law takes place.

In order to overcome this problem, in Ref. \cite{zeno1} a definition has been
introduced which does not depend on the Taylor expansion of the function
$p(t).$ The time $\tau_{M}$ is defined as the instant of time at which the
deviation of the function $p(t)$ from the exponential behavior $e^{-\Gamma t}$
is maximal:%
\begin{equation}
\max\left(  p(t)-e^{-\Gamma t}\right)  \rightarrow t=\tau_{M}\text{ .}
\label{taum}%
\end{equation}
Clearly
\begin{equation}
\frac{d}{dt}\left(  p(t)-e^{-\Gamma t}\right)  _{t=\tau_{M}}=0\text{ ;}%
\end{equation}
in all practical cases $\tau_{M}$ corresponds to the first zero of the
derivative of the function $p(t)-e^{-\Gamma t}$. The function $p(t)-e^{-\Gamma
t}$ has been plotted in Fig. 3 for the numerical values of Eq. (\ref{numval}).
The maximum is realized for $\tau_{M}=1.18\left[  2m_{2}\right]  ^{-1}$,
implying the ratio $\tau_{M}/\tau_{\text{BW}}=0.26.$ In the present model, the
non-exponential behavior elapses for $\sim2\tau_{M},$ meaning that in this
example is about 50\% of the tree-level lifetime of the unstable particle.

Further comments are in order:

(i) In the present work we consider field theoretical examples with some
finite value of the cutoff. It is indeed a natural requirement that at some
energy $x\gtrsim\Lambda$ the loop function $\Sigma(x^{2},m^{2})$ is
suppressed. The actual value of the constant $\Lambda$ depends on the theory
under study. For instance, working with hadronic theories implies $\Lambda
\sim1$-$2$ GeV \cite{amslerrev}. On the contrary, for elementary Standard
Model processes, the cutoff $\Lambda$ is of the order of $M_{\text{Planck}}$
(alternatives are $\Lambda\sim M_{GUT}\sim10^{16}$ GeV and, in some recent
scenarios with large extradimension(s), even smaller with $\Lambda\gtrsim1$ TeV).

(ii) One can also regard the theory of Eq. (\ref{lag}) as a
(super)renormalizable toy model with a very large value of the cutoff, such as
$\Lambda\sim M_{\text{Planck}}\sim10^{19}$ GeV. In this case the first
derivative vanishes, $p^{\prime}(0)=0$, and all the other derivatives,
including the second derivative $p^{\prime\prime}(0),$ are large but finite:
for instance, the quantity $p^{\prime\prime}(0)$ is proportional to
$\ln\Lambda$. This is easily understandable, being the asymptotic behavior for
large $x$ given by $d_{S}(x)\sim x^{-3}$, thus implying that $\left\langle
x^{2}\right\rangle =\int_{-\infty}^{\infty}x^{2}d_{S}(x)\mathrm{dx}\sim
\ln\Lambda$. In the present superrenormalizable case, the functions $p(t)$ for
$\Lambda\sim1$-$2$ GeV and for $\Lambda\sim M_{\text{Planck}}$ differ only
slightly: the deviations from the exponential law last for a similar amount of
time and the value of the ratio $\tau_{M}/\tau_{\text{BW}}$ shows a very weak
dependence on the cutoff \cite{zeno1}. This example shows that, in the case of
large $\Lambda$ (as for instance, $\Lambda\sim M_{\text{Planck}}$), although
the Zeno time $\tau_{Z}\sim1/\sqrt{\ln\Lambda}$ turns out to be very small (as
it was obtained in Refs. \cite{alvarez,maianitesta} in the framework of
perturbation theory), $\tau_{Z}$ offers only an underestimation of the actual
non-exponential time interval, which is sizable also for large values of the cutoff.

(iii) In the nonrelativistic Breit-Wigner limit of Eq. (\ref{dbw}) the first
derivative $p^{\prime}(0)$ is --strictly speaking-- not defined. However, by
using the fact that the amplitude reads $p_{\text{BW}}(t>0)=e^{-\Gamma t}$,
one obviously has $\lim_{t\rightarrow0^{+}}p_{\text{BW}}^{\prime}(t)=-\Gamma$.
In other words, the function $p_{\text{BW}}^{\prime}(t)$ can be easily
continued to the point $t=0$.

(iv) The present results are based on the (resummed) 1-loop approximation. The
inclusion of higher order, such as the sunset diagram in which a particle $S$
is exchanged by the two particles $\varphi$ circulating in the loop of Fig.
1.b, is a task for the future. It is expected that higher order terms do not
change substantially the results: in fact, higher order diagrams are
suppressed by the multiplication of vertex functions. (Moreover, in the
framework of hadronic theories, higher order are suppressed in the so-called
large-$N_{c}$ approximation \cite{witten}. The behavior of a theory similar to
that of Eq. (\ref{lag}) in the large-$N_{c}$ limit is discussed in Ref.
\cite{dynrec}).

(v) The resummation of the loop diagram in Fig. 1.b, expressed in its
unregularized form in Eq. (\ref{self}) and in its regularized form in Eq.
(\ref{luporeg}) reported in the Appendix A, has been a crucial step toward the
presented results, also in connection with the correct normalization of $p(t)$
and the vanishing of $p^{\prime}(0)$ (necessary to show the non-exponential
behavior). If, instead, one would consider only the perturbative expressions
of the propagator at the 1-loop ($g_{i}^{2})$ level, one obtains
\begin{equation}
G_{S}^{\text{pert}}(p^{2})=\frac{1}{p^{2}-M_{0}^{2}+i\varepsilon}+\left(
\frac{1}{p^{2}-M_{0}^{2}+i\varepsilon}\right)  ^{2}\left[  (\sqrt{2}g_{1}%
)^{2}\Sigma(p^{2},m_{1}^{2})+(\sqrt{2}g_{2})^{2}\Sigma(p^{2},m_{2}%
^{2})\right]  +...\text{ ,}%
\end{equation}
thus the spectral function in this case would take the form%
\begin{equation}
d_{S}^{\text{pert}}(x)=\frac{2x}{\pi}\operatorname{Im}G_{S}^{\text{pert}%
}(p^{2}=x^{2})=\delta(x-M_{0})+\frac{4g_{1}^{2}}{\pi}\frac{\sqrt{\frac{x^{2}%
}{4}-m_{1}^{2}}}{(x^{2}-M_{0}^{2})^{2}}+\frac{4g_{2}^{2}}{\pi}\frac
{\sqrt{\frac{x^{2}}{4}-m_{2}^{2}}}{(x^{2}-M_{0}^{2})^{2}}+...
\end{equation}
The normalization of $d_{S}^{\text{pert}}(x)$ is not anymore explicitly
fulfilled for each value of $g$ and an unphysical singularity for the bare
value of the mass $x=M_{0}$ is present. (This infinity is removed when the
resummation is performed.) It would be cumbersome to derive analytical results
with this perturbative expressions; moreover, a numerical study could hardly
be performed, thus showing the necessity of the resummation \emph{before} the
temporal analysis is investigated.

(vi) The issue of the preparation of the state $\left\vert S\right\rangle $ at
the time $t=0$ is well defined when the state is long lived, but more
controversial for short living particles. A related difficulty is the
(in)dependence on the representation choice \cite{Kamefuchi:1961sb}, see also
the corresponding discussions in Ref. \cite{zeno1}. Here it suffices to say
that the change of the representation of the fields is equivalent to the
change of the initial state. Nevertheless, also in this case deviations from
the exponential law take place, but the mass distribution would have a
different form.

(vii) As a last step we discuss the validity of the employed model of Eq.
(\ref{lag}) upon variation of the parameters. The choice of the numerical
values in\ Eq. (\ref{numval}) has been done in order to obtain figures which
allow for a clear presentation of the underlying equations. However, the
described patterns take place also when the parameters are varied: in
particular, when the coupling constants $g_{1}$ and $g_{2}$ are decreased, the
time interval in which sizable variations from the exponential decay law take
place becomes smaller. In the opposite direction, when the coupling constants
$g_{1}$ and $g_{2}$ become larger, this time interval grows. However, care is
needed when increasing the values of the coupling constants: beyond a certain
threshold, the validity of the K\"{a}llen-Lehman representation, and thus of
the normalization condition of Eq. (\ref{norm}), does not any longer hold, see
the detailed discussion in Ref. \cite{achasov}. Note also that the QFT model
of Eq. (\ref{lag}) does not show, for any value of the parameters,
repopulation effect. In fact, the function $h(t)$ is always positive in this model.

\subsection{The quantum Zeno effect}

It is interesting and amusing to briefly review the so-called quantum Zeno
effect. To this end we define the function $\gamma(t)$ as in Ref.
\cite{facchiprl}:%
\begin{equation}
\gamma(t)=-\frac{\ln p(t)}{t}\leftrightarrow p(t)=e^{-\gamma(t)t}\text{ .}%
\end{equation}
In Fig. 5 the function $\gamma(t)$ is plotted and is compared to the
tree-level decay width $\Gamma$. It is visible that for short times
$\gamma(t)$ is considerably smaller than $\Gamma$, see the discussion below.

Let us now suppose to perform a single measurement of the system at the time
$\tau>0.$ The probability that the unstable state $S$ did not decay at the
time $\tau$ reads%
\begin{equation}
p_{\tau}=p(\tau)=e^{-\gamma(\tau)\tau}\simeq e^{-\Gamma\tau}%
\end{equation}
where in the last passage we assumed that the time $\tau$ is large enough in
order that $\gamma(\tau)\simeq\Gamma.$

Instead, let us suppose to perform $N$ measurements at the time intervals
$t_{\ast}$, such that $\tau=Nt_{\ast}$. The probability that at the time
$\tau$ (after the $N$ measurements) the state did not decay is given by%
\begin{equation}
p_{N\text{-}t_{\ast}}=\left(  e^{-\gamma(t_{\ast})t_{\ast}}\right)
^{N}=e^{-\gamma(t_{\ast})Nt_{\ast}}=e^{-\gamma(t_{\ast})\tau}\text{ .}%
\end{equation}
When $t_{\ast}$ belongs to the `non-exponential' time-interval, $0\leq
t_{\ast}\lesssim\tau_{M},$ one has $\gamma(t_{\ast})<\Gamma.$ Indeed,
considering that the function $\gamma(t)$ vanishes for $t=0$,
\begin{equation}
\gamma(0)=\left(  -\frac{p^{\prime}(t)}{p(t)}\right)  _{t=0}=-p^{\prime
}(0)=0\text{ },
\end{equation}
one can make the quantity $\gamma(t_{\ast})$ as small as desired by choosing
$t_{\ast}$ $\ll\tau_{M}.$

In general, for $0\leq t_{\ast}\lesssim2\tau_{M},$ the inequality
\begin{equation}
p_{N\text{-}t_{\ast}}=e^{-\gamma(t_{\ast})\tau}>p_{\tau}\text{ }\simeq
e^{-\Gamma\tau}\text{.}%
\end{equation}
holds. Its meaning is clear but astonishing: the subsequent measurements of
the unstable particle $S$ increased the probability that it did not decay.
(Formally, one performs a `measurement' of the unperturbed Hamiltonian $H_{0}%
$, in which the coupling constants $g_{i}$ are set to zero). This property has
been indeed verified experimentally by using ions trapped in a potential
barrier \cite{raizen2}.

For $t_{\ast}\rightarrow0,$ i.e. for $N\rightarrow\infty$ (a situation which
corresponds to a `continuos' measurement) one obtains that $p_{N\text{-}%
t_{\ast}}\rightarrow1.$ The unstable particle does not decay at all! This is
the famous quantum Zeno paradox.%

\begin{figure}
[ptb]
\begin{center}
\includegraphics[
height=2.7735in,
width=4.6293in
]%
{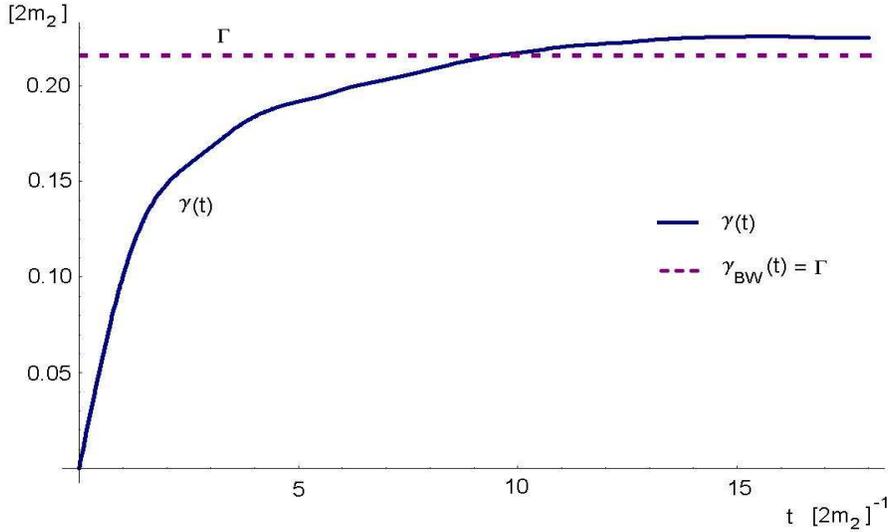}%
\caption{The function $\gamma(t)=-\left[  \ln p(t)\right]  /t$ is plotted. For
comparison, the constant BW limit $\gamma_{\text{BW}}(t)=-\left[  \ln
p_{\text{BW}}(t)\right]  /t=\Gamma$ is also shown. For short times
$\gamma(t)<\Gamma,$ rendering the quantum Zeno effect possible. As usual, the
numerical values of Eq. (\ref{numval}) have been used.}%
\end{center}
\end{figure}

As Fig. 5 shows, there are also time intervals in which $\gamma(t)>\Gamma.$
This implies that for $t_{\ast}$ satisfying this relation a faster decay rate
is obtained when the state of the unstable particle is observed at intervals
of $t_{\ast}$. This situation corresponds to the so called quantum Anti-Zeno
effect \cite{zheng,facchiprl}.

We conclude this brief recall of the Zeno effect by noticing that it does not
take place in the Breit-Wigner limit. In fact, in this limit one has
$\gamma(t)=\Gamma$ for each $t$, thus meaning that $p_{N-t_{\ast}}=p_{\tau},$
independently on the value of $t_{\ast}$ and $\tau$. Making a single
measurements or many intermediate measurements does not lead to any change in
the survival probability.

\section{General properties of decays in QM: Lee Hamiltonian and link to QFT}

\subsection{QM with Lee Hamiltonian(s)}

We now study a similar system in the framework of quantum mechanics. In order
to obtain irreversible dynamics in a QM approach, one needs to couple the
unstable state $\left\vert S\right\rangle $ to a continuum of states. The
formalism described here relies on the so-called Lee Hamiltonians \cite{lee}
(later on denoted also as QM-LH approach) and is a generalization to the
two-channel case of the work in Ref. \cite{facchiprl}. In this section we
recall general properties of the LH approach with the special aim of showing
the similarities with the genuine QFT approach presented in the previous
section. In this context we also refer to Ref. \cite{chiu}, where a solvable
Lee Hamiltonian has been studied in great detail and the relation to QFT has
been described.

For transparency, we first present the Hamiltonian using a discrete sum and
later on perform the transition to the continuous results. The Hamiltonian
under consideration is given by%

\begin{equation}
H=H_{0}+H_{int}\text{ , }H_{int}=H_{int}^{(1)}+H_{int}^{(2)}\text{ ,}
\label{hleenr}%
\end{equation}
where%
\begin{equation}
H_{0}=M_{0}\left\vert S\right\rangle \left\langle S\right\vert +\sum
_{\mathbf{k}}\omega_{1}(\mathbf{k})\left\vert 1,\mathbf{k}\right\rangle
\left\langle 1,\mathbf{k}\right\vert +\sum_{\mathbf{k}}\omega_{2}%
(\mathbf{k})\left\vert 2,\mathbf{k}\right\rangle \left\langle 2,\mathbf{k}%
\right\vert \text{ ,}%
\end{equation}

\begin{align}
H_{int}^{(1)}  &  =\sum_{\mathbf{k}}g_{1}\frac{f_{1}(\mathbf{k})}{\sqrt{V}%
}\left(  \left\vert 1,\mathbf{k}\right\rangle \left\langle S\right\vert
+\left\vert S\right\rangle \left\langle 1,\mathbf{k}\right\vert \right)
\text{ ,}\label{mix1}\\
H_{int}^{(2)}  &  =\sum_{\mathbf{k}}g_{2}\frac{f_{2}(\mathbf{k})}{\sqrt{V}%
}\left(  \left\vert 2,\mathbf{k}\right\rangle \left\langle S\right\vert
+\left\vert S\right\rangle \left\langle 2,\mathbf{k}\right\vert \right)
\text{ .} \label{mix2}%
\end{align}
In order to establish a contact with the study of Sec. 2, the variable
$\mathbf{k}$ stays for $\mathbf{k}=2\pi\mathbf{n}/L$ and $V=L^{3}$. The
analogy is clear: the state $\left\vert S\right\rangle $ corresponds to the
particle $S$ of Eq. (\ref{lag}) and the states $\left\vert 1,\mathbf{k}%
\right\rangle $, $\left\vert 2,\mathbf{k}\right\rangle $ are analogous to the
two-particles states $\left\vert \varphi_{1,\mathbf{k}}\varphi_{1,-\mathbf{k}%
}\right\rangle $ and $\left\vert \varphi_{2,\mathbf{k}}\varphi_{2,-\mathbf{k}%
}\right\rangle $ of Sec. 2. The mixing terms in Eqs. (\ref{mix1})-(\ref{mix2})
correspond to the interaction terms $g_{1}S\varphi_{1}^{2}$ and $g_{2}%
S\varphi_{2}^{2}$ of Eq. (\ref{lag}), respectively.

The transition to the continuous limit is obtained for%
\begin{equation}
\sum_{\mathbf{k}}\rightarrow V\int\frac{d^{3}k}{(2\pi)^{3}}\text{ ,
}\left\vert i,\mathbf{k}\right\rangle \rightarrow\sqrt{\frac{(2\pi)^{3}}{V}%
}\left\vert i,\mathbf{k}\right\rangle \text{ , }i=1,2.
\end{equation}

The time-evolution operator is given $U(t)=e^{-iHt}$ can be expressed in terms
of a Fourier transform:%
\begin{equation}
U(t)=e^{-iHt}=\frac{i}{2\pi}\int_{-\infty}^{+\infty}dE\frac{1}%
{E-H+i\varepsilon}e^{-iEt}\text{ }=\frac{i}{2\pi}\int_{-\infty}^{+\infty
}dEG(E)e^{-iEt}\text{ ,} \label{u(t)}%
\end{equation}
where the operator
\begin{equation}
G(E)=\frac{1}{E-H+i\varepsilon}%
\end{equation}
has been introduced. One can rewrite $G(E)$ as follow:%
\begin{equation}
G(E)=\frac{1}{E-H+i\varepsilon}=\frac{1}{E-H_{0}+i\varepsilon}\frac{1}%
{1-\frac{H_{int}}{E-H_{0}+i\varepsilon}}=\frac{1}{E-H_{0}+i\varepsilon}%
\sum_{n=0}^{\infty}\left(  H_{int}\frac{1}{E-H_{0}+i\varepsilon}\right)  ^{n}
\label{sumg}%
\end{equation}
The propagator $G_{S}(E)$ of the unstable state $\left\vert S\right\rangle $
is defined as%
\begin{equation}
G_{S}(E)=\left\langle S\right\vert G(E)\left\vert S\right\rangle \text{ }%
\end{equation}
and can be evaluated using Eq. (\ref{sumg}):
\begin{equation}
G_{S}(E)=\frac{1}{E-M_{0}+i\varepsilon}\sum_{n=0}^{\infty}\left\langle
S\right\vert \left(  H_{int}\frac{1}{E-H_{0}+i\varepsilon}\right)
^{n}\left\vert S\right\rangle
\end{equation}
Explicitly, the cases $n=1$ and $n=2$ read:%
\begin{equation}
n=1\rightarrow\frac{1}{E-M_{0}+i\varepsilon}\left\langle S\right\vert
H_{int}\left\vert S\right\rangle =0\text{ ,}%
\end{equation}

\begin{equation}
n=2\rightarrow\left\langle S\right\vert \left(  H_{int}\frac{1}{E-H_{0}%
+i\varepsilon}\right)  ^{2}\left\vert S\right\rangle =\left\langle
S\right\vert H_{int}\frac{1}{E-H_{0}+i\varepsilon}H_{int}\left\vert
S\right\rangle \frac{1}{E-M_{0}+i\varepsilon}\text{ .}%
\end{equation}
The recursive quantity found for each $n$ even is $\Pi(E)$:%
\begin{align}
\Pi(E)  &  =-\left\langle S\right\vert H_{int}\frac{1}{E-H_{0}+i\varepsilon
}H_{int}\left\vert S\right\rangle =\\
-  &  \left\langle S\right\vert H_{int}\left(  \left\vert 1,\mathbf{k}%
\right\rangle \left\langle 1,\mathbf{k}\right\vert +\left\vert 2,\mathbf{k}%
\right\rangle \left\langle 2,\mathbf{k}\right\vert \right)  \frac{1}%
{E-H_{0}+i\varepsilon}\left(  \left\vert 1,\mathbf{q}\right\rangle
\left\langle 1,\mathbf{q}\right\vert +\left\vert 2,\mathbf{q}\right\rangle
\left\langle 2,\mathbf{q}\right\vert \right)  H_{int}\left\vert S\right\rangle
\text{ ,}%
\end{align}
where the sums over $\mathbf{k}$ and $\mathbf{q}$ are understood. Using Eqs.
(\ref{mix1})-(\ref{mix2}) and the orthonormality of the basis we get:%
\begin{equation}
\Pi(E)=-\sum_{\mathbf{k}}\frac{1}{V}\frac{g_{1}^{2}f_{1}^{2}(\mathbf{k}%
)}{E-\omega_{1}(\mathbf{k})+i\varepsilon}-\sum_{\mathbf{k}}\frac{1}{V}%
\frac{g_{2}^{2}f_{2}^{2}(\mathbf{k})}{E-\omega_{2}(\mathbf{k})+i\varepsilon
}\text{ .}%
\end{equation}
In the continuous limit the basic quantity $\Pi(E)$ becomes:%
\begin{equation}
\Pi(E)=-\int\frac{d^{3}k}{(2\pi)^{3}}\frac{g_{1}^{2}f_{1}^{2}(\mathbf{k}%
)}{E-\omega_{1}(\mathbf{k})+i\varepsilon}-\int\frac{d^{3}k}{(2\pi)^{3}}%
\frac{g_{2}^{2}f_{2}^{2}(\mathbf{k})}{E-\omega_{2}(\mathbf{k})+i\varepsilon
}=g_{1}^{2}\Sigma_{1}(E)+g_{2}^{2}\Sigma_{2}(E)\text{ .} \label{delta}%
\end{equation}

The propagator $G_{S}(E)$ of the state $\left\vert S\right\rangle $ reads then%
\begin{equation}
G_{S}(E)=\left\langle S\right\vert G(E)\left\vert S\right\rangle =\frac
{1}{E-M_{0}+\Pi(E)+i\varepsilon}\text{ .}%
\end{equation}
The similarities between the present QM-LH approach and the QFT approach
described in the previous section are clear (see for instance the form of the
propagator in Eq. (\ref{propqft})), the main difference being the emergence of
nonrelativistic propagator in the present case. The quantity $\Pi(E)$
represents the analogous of the loop contributions.

We can now calculate the survival amplitude $a(t)=\left\langle S\right\vert
U(t)\left\vert S\right\rangle $:
\begin{equation}
a(t)=\left\langle S\right\vert U(t)\left\vert S\right\rangle =\left\langle
S\right\vert \frac{i}{2\pi}\int_{-\infty}^{+\infty}dE\frac{1}{E-H+i\varepsilon
}e^{-iEt}\left\vert S\right\rangle =\frac{i}{2\pi}\int_{-\infty}^{+\infty
}dEG_{S}(E)e^{-iEt}\text{.}%
\end{equation}
Using the dispersion relation%
\begin{equation}
G_{S}(E)=\frac{1}{\pi}\int_{-\infty}^{\infty}dx\frac{\operatorname{Im}%
[G_{S}(x)]}{x-E-i\varepsilon} \label{disprel}%
\end{equation}
the survival amplitude $a(t)$ can be expressed as%
\begin{equation}
a(t)=\int_{-\infty}^{\infty}\mathrm{dx}d_{S}(x)e^{-ixt}%
\end{equation}
where in this context the identification
\begin{equation}
d_{S}(x)=\frac{1}{\pi}\left\vert \operatorname{Im}[G_{S}(x)]\right\vert
\label{dsqm}%
\end{equation}
has been made. Again, the expression (\ref{dsqm}) and its QFT-counterpart of
Eq. (\ref{ds}) are formally equivalent (a part from unimportant
normalizations). Obviously, all the discussion about the deviations from the
exponential law, the Breit-Wigner limit, the vanishing of $p^{\prime}(0)$ and
the quantum Zeno effect can be easily repeated in the present QM-LH context.

\subsection{Relativistic generalization of the LH and link to QFT}

In order to obtain expressions which are equal to those of QFT of Sec. 2, one
has to consider a slight modification of the Lee model. The issue is that in
QFT one has particles with positive energy propagating forward in time and
particles with negative energy propagating backwards in time (i.e.
antiparticles). In our QFT model of Eq. (\ref{lag}) particles and
antiparticles coincide, but it is anyhow the combination of both contributions
which give rise to the usual relativistic propagator of the form
$1/(p^{2}-m^{2}+i\varepsilon)$.

The QFT expressions can be obtained in a LH approach by modifying the time
evolution operator (\ref{u(t)}) in the following way:%

\begin{align}
U(t)  &  =\frac{i}{2\pi}\int_{-\infty}^{+\infty}dE\left(  \frac{1}%
{E-H+i\varepsilon}+\frac{1}{E+H-i\varepsilon}\right)  e^{-iEt}\\
&  =\frac{i}{2\pi}\int_{-\infty}^{+\infty}dE\frac{2E}{E^{2}-H^{2}%
+i\varepsilon}e^{-iEt}\text{ }=\frac{i}{2\pi}\int_{-\infty}^{+\infty
}dE2EG(E)e^{-iEt}\text{,}%
\end{align}
where the operator $G(E)$ reads now
\begin{equation}
G(E)=\frac{1}{E^{2}-H^{2}+i\varepsilon}\text{ .}%
\end{equation}
The integral has been done according to the famous Feynman prescription. If,
as a simple example, one considers a free particle $\left\vert S\right\rangle
$ for which $H_{0}^{2}=M_{0}^{2}\left\vert S\right\rangle \left\langle
S\right\vert $, one finds the usual propagator form (in the rest frame of $S$)
$G_{S}(E)=\left\langle S\right\vert G(E)\left\vert S\right\rangle
=1/(E^{2}-M_{0}^{2}+i\varepsilon)$.

We consider a model in which we can write $H^{2}$ as%
\begin{equation}
H^{2}=H_{0}^{2}+H_{int}^{2},\text{ }H_{int}^{2}=H_{int}^{(1)2}+H_{int}^{(2)2}%
\end{equation}
where
\begin{equation}
H_{0}^{2}=M_{0}^{2}\left\vert S\right\rangle \left\langle S\right\vert
+\sum_{\mathbf{k}}\omega_{1}^{2}(\mathbf{k})\left\vert 1,\mathbf{k}%
\right\rangle \left\langle 1,\mathbf{k}\right\vert +\sum_{\mathbf{k}}%
\omega_{2}^{2}(\mathbf{k})\left\vert 2,\mathbf{k}\right\rangle \left\langle
2,\mathbf{k}\right\vert \text{ ,}%
\end{equation}%
\begin{align}
H_{int}^{(1)2}  &  =\sum_{\mathbf{k}}g_{1}\frac{f_{1}(\mathbf{k})}{\sqrt{V}%
}\left(  \left\vert 1,\mathbf{k}\right\rangle \left\langle S\right\vert
+\left\vert S\right\rangle \left\langle 1,\mathbf{k}\right\vert \right)
\text{ ,}\\
H_{int}^{(2)2}  &  =\sum_{\mathbf{k}}g_{2}\frac{f_{2}(\mathbf{k})}{\sqrt{V}%
}\left(  \left\vert 2,\mathbf{k}\right\rangle \left\langle S\right\vert
+\left\vert S\right\rangle \left\langle 2,\mathbf{k}\right\vert \right)
\text{ .}%
\end{align}
We then follow the same steps of Sec. 3.1 and introduce the propagator as%
\begin{equation}
G_{S}(E)=\left\langle S\right\vert G(E)\left\vert S\right\rangle =\frac
{1}{E^{2}-M_{0}^{2}+\Pi(E)+i\varepsilon}\text{ ,}%
\end{equation}
where $\Pi(E)$ is calculated in the same way as in Sec. 3.1 and --as
expected-- turns out to be the relativistic generalization of Eq.
(\ref{delta}):
\begin{equation}
\Pi(E)=-\int\frac{d^{3}k}{(2\pi)^{3}}\frac{g_{1}^{2}f_{1}^{2}(\mathbf{k}%
)}{E^{2}-\omega_{1}^{2}(\mathbf{k})+i\varepsilon}-\int\frac{d^{3}k}{(2\pi
)^{3}}\frac{g_{2}^{2}f_{2}^{2}(\mathbf{k})}{E^{2}-\omega_{2}^{2}%
(\mathbf{k})+i\varepsilon}=g_{1}^{2}\Sigma_{1}(E)+g_{2}^{2}\Sigma_{2}(E)\text{
.}%
\end{equation}
Notice that the propagator $G_{S}(E)=$ $\left\langle S\right\vert
G(E)\left\vert S\right\rangle $ takes a form which is formally equivalent to
Eq. (\ref{propqft}) obtained in QFT.

In particular, when we make the identifications
\begin{equation}
\omega_{1}^{2}(\mathbf{k})=4(\mathbf{k}^{2}+m_{1}^{2})\text{ , }\omega_{2}%
^{2}(\mathbf{k})=4(\mathbf{k}^{2}+m_{2}^{2})\text{ ,} \label{eq1}%
\end{equation}
and%
\begin{equation}
f_{1}(\mathbf{k})=\sqrt{2}\frac{\tilde{\phi}(\mathbf{k})}{\left(
\mathbf{k}^{2}+m_{1}^{2}\right)  ^{1/4}}\text{ , }f_{2}(\mathbf{k})=\sqrt
{2}\frac{\tilde{\phi}(\mathbf{k})}{\left(  \mathbf{k}^{2}+m_{2}^{2}\right)
^{1/4}}\text{ ,} \label{eq2}%
\end{equation}
we get indeed in the present Lee model the \emph{same }mathematical expression
for the propagator of the QFT approach of Sec. 2. Note, the additional factor
$\sqrt{2}$ is here necessary to make contact with the identical particles in
QFT and the cutoff function $\tilde{\phi}(\mathbf{q})$ is described in detail
in Appendix A and takes the form $\tilde{\phi}(\mathbf{q})=\theta(\Lambda
^{2}-\mathbf{q}^{2})$ for the numerical evaluations presented in Figs. 2-10.
Loosely speaking, the QFT case corresponds to a particular choice of the
functions $\omega_{i}(\mathbf{k})$ and $f_{i}(\mathbf{k})$ of the Lee
Hamiltonian studied here. This is an interesting property which we will use in
Sec. 5.

By making use of dispersion relations we arrive at the usual expression for
the survival amplitude $a(t)$:%

\begin{equation}
a(t)=\int_{-\infty}^{\infty}\mathrm{dx}d_{S}(x)e^{-ixt}\text{ ,}%
\end{equation}
where
\begin{equation}
d_{S}(x)=\frac{2x}{\pi}\left\vert \lim_{\varepsilon\rightarrow0}%
\operatorname{Im}[\left\langle S\right\vert G_{S}(x)\left\vert S\right\rangle
]\right\vert
\end{equation}
is identical to Eq. (\ref{ds}). This shows once more the tight connection of
the present formalism with that of Sec. 2.

\section{Two-channel case: quantum field theory}

\subsection{The decay probabilities $w_{1}(t)$ and $w_{2}(t)$ and the
corresponding densities $h_{1}(t)$ and $h_{2}(t)$: the formal expressions}

In this section we come back to the quantum field theoretical model of Eq.
(\ref{lag}). The central quantities under investigation are the decay
probability densities $h_{1}(t)$ and $h_{2}(t)$ and the corresponding
integrated decay probabilities $w_{1}(t)$ and $w_{2}(t)$. The object
$h_{1}(t)dt$ represents the probability that the particle $S$ decays in the
first channel ($S\rightarrow\varphi_{1}\varphi_{1}$) in the time interval
between $t$ and $t+dt$; as a consequence, the integral $w_{1}(t)=\int_{0}%
^{t}\mathrm{du}h_{1}(u)$ represents the probability that the particle $S$
decays into the first channel in the time interval $(0,t)$. The quantity
$w_{1}(t)$ can be formally expressed as the sum of the probabilities of decays
into $\varphi_{1}\varphi_{1}$ for different values of the three-momentum
$\mathbf{k}$:
\begin{equation}
w_{1}(t)=\int_{0}^{t}\mathrm{du}h_{1}(u)=\sum_{\mathbf{k}}\left\vert
\left\langle \varphi_{1,\mathbf{k}}\varphi_{1,-\mathbf{k}}\left\vert
e^{-iHt}\right\vert S\right\rangle \right\vert ^{2}\text{ .} \label{w1}%
\end{equation}

Similarly, $h_{2}(t)dt$ represents the probability that the particle $S$
decays in the second channel ($S\rightarrow\varphi_{2}\varphi_{2}$) in the
time interval between $t$ and $t+dt$ and the integral%
\begin{equation}
w_{2}(t)=\int_{0}^{t}\mathrm{du}h_{2}(u)=\sum_{\mathbf{k}}\left\vert
\left\langle \varphi_{2,\mathbf{k}}\varphi_{2,-\mathbf{k}}\left\vert
e^{-iHt}\right\vert S\right\rangle \right\vert ^{2}\text{ } \label{w2}%
\end{equation}
represents the probability that the particle $S$ decays into the second
channel in the time interval $(0,t)$. The equality $w(t)=w_{1}(t)+w_{2}(t)$
holds by construction (see Eq. (\ref{intothsv})) and is intuitively clear: the
total decay probability between $(0,t)$ is the sum of the decay probabilities
in each channel in the same time interval.

The symbolic expressions for $h_{1}(t)$ and $h_{2}(t)$ in terms of state
vectors are obtained by deriving Eqs. (\ref{w1}) and (\ref{w2}):
\begin{align}
h_{1}(t)  &  =\frac{d}{dt}\left(  \sum_{\mathbf{k}}\left\vert \left\langle
\varphi_{1,\mathbf{k}}\varphi_{1,-\mathbf{k}}\left\vert e^{-iHt}\right\vert
S\right\rangle \right\vert ^{2}\right)  \text{ ,}\label{h1form}\\
h_{2}(t)  &  =\frac{d}{dt}\left(  \sum_{\mathbf{k}}\left\vert \left\langle
\varphi_{2,\mathbf{k}}\varphi_{2,-\mathbf{k}}\left\vert e^{-iHt}\right\vert
S\right\rangle \right\vert ^{2}\right)  \text{ .} \label{h2form}%
\end{align}
In this section the decay probability densities $h_{1}(t)$ and $h_{2}(t)$ are
not calculated directly, because the evaluation of the matrix elements
$\left\langle \varphi_{1,\mathbf{k}}\varphi_{1,-\mathbf{k}}\left\vert
e^{-iHt}\right\vert S\right\rangle $ and the subsequent partial sum is not an
easy task in QFT. We will follow a different way by starting from the decay
probability density $h(t)$, which by construction is the sum of $h_{1}(t)$ and
$h_{2}(t)$. It is then possible to postulate the expressions for the latter
quantities and then a posteriori verify that they fulfill all the required
properties. We will check the numerical validity of our solutions with the
help of the QM-LH approach (both in the nonrelativistic and the relativistic
frameworks) in Sec. 5.

An important object of interest in this section is the ratio of decay
probability densities $R(t)$ defined as
\begin{equation}
R(t)=\frac{h_{1}(t)}{h_{2}(t)}\text{ }. \label{r}%
\end{equation}
It represents the ratio of the probabilities that the particle $S$ decays in
the first channel and in the second channel in the time interval between $t$
and $t+dt$. Its study is interesting because, as we shall see, $R(t)$ shows a
peculiar behavior as function of time.

For a better understanding of the presented functions several plots for the
numerical values of Eq. (\ref{numval}) are presented.

\subsection{The partial distributions $d_{S}^{(1)}(x)$ and $d_{S}^{(2)}(x)$
and amplitudes $a_{1}(t)$ and $a_{2}(t)$}

The distribution $d_{S}(x)$ can be written as the sum of the two functions
$d_{S}^{(1)}(x)$ and $d_{S}^{(2)}(x)$:%
\begin{equation}
d_{S}(x)=d_{S}^{(1)}(x)+d_{S}^{(2)}(x)\text{ ,}%
\end{equation}
with%
\begin{equation}
d_{S}^{(1)}(x)=\frac{2x}{\pi}\lim_{\varepsilon\rightarrow0}\frac{2g_{1}%
^{2}\operatorname{Im}\left[  \Sigma(x^{2},m_{1}^{2})\right]  +\varepsilon
}{\left(  x^{2}-M_{0}^{2}+\operatorname{Re}\Pi(x^{2})\right)  ^{2}+\left(
\operatorname{Im}\Pi(x^{2})+\varepsilon\right)  ^{2}}\text{ ,} \label{ds1}%
\end{equation}%
\begin{equation}
d_{S}^{(2)}(x)=\frac{2x}{\pi}\lim_{\varepsilon\rightarrow0}\frac{2g_{2}%
^{2}\operatorname{Im}\left[  \Sigma(x^{2},m_{2}^{2})\right]  +\varepsilon
}{\left(  x^{2}-M_{0}^{2}+\operatorname{Re}\Pi(x^{2})\right)  ^{2}+\left(
\operatorname{Im}\Pi(x^{2})+\varepsilon\right)  ^{2}}\text{ .} \label{ds2}%
\end{equation}
The integrals
\begin{equation}
\int_{0}^{\infty}\mathrm{dx}d_{S}^{(1)}(x)\text{ and }\int_{0}^{\infty
}\mathrm{dx}d_{S}^{(2)}(x)
\end{equation}
represent the decay fractions, i.e. the probabilities that the particle $S$
decays in channel $1$ or in channel $2$ in the time interval $(0,\infty)$,
respectively \cite{lupoder}. Obviously, their sum is unity in virtue of Eq.
(\ref{norm}). Note, in the BW limit the integral $\int_{0}^{\infty}%
\mathrm{dx}d_{S}^{(i)}(x)$ represents the ratio $\Gamma_{i}/\Gamma,$ see Sec.
4.3.5 for details.

The functions $d_{S}^{(1)}(x)$ and $d_{S}^{(2)}(x)$ have been plotted in Fig.
6 for the numerical values of Eq. (\ref{numval}).

We now introduce the Fourier transforms of $d_{S}^{(1)}(x)$ and $d_{S}%
^{(2)}(x)$:%

\begin{equation}
a_{1}(t)=\int_{-\infty}^{\infty}\mathrm{dx}d_{S}^{(1)}(x)e^{-ixt}\text{ ,}%
\end{equation}

\begin{equation}
a_{2}(t)=\int_{-\infty}^{\infty}\mathrm{dx}d_{S}^{(2)}(x)e^{-ixt}\text{ .}%
\end{equation}
The survival amplitude $a(t)$ can be rewritten as%
\begin{equation}
a(t)=\int_{-\infty}^{\infty}\mathrm{dx}d_{S}(x)e^{-ixt}=\int_{-\infty}%
^{\infty}\mathrm{dx}d_{S}^{(1)}(x)e^{-ixt}+\int_{-\infty}^{\infty}%
\mathrm{dx}d_{S}^{(2)}(x)e^{-ixt}=a_{1}(t)+a_{2}(t)\text{ .}%
\end{equation}
Then, the survival probability $p(t)$ can be expressed as:%
\begin{align}
p(t)  &  =a^{\ast}(t)a(t)=\left(  a_{1}^{\ast}(t)+a_{2}^{\ast}(t)\right)
\left(  a_{1}(t)+a_{2}(t)\right) \nonumber\\
&  =a_{1}^{\ast}(t)a_{1}(t)+a_{1}^{\ast}(t)a_{2}(t)+a_{1}(t)a_{2}^{\ast
}(t)+a_{2}^{\ast}(t)a_{2}(t)\text{ }\nonumber\\
&  =A_{1}(t)+2A_{mix}(t)+A_{2}(t)\text{ ,}%
\end{align}
where the three real quantities
\begin{align}
A_{1}(t)  &  =\left\vert a_{1}(t)\right\vert ^{2}=\left\vert \int_{-\infty
}^{\infty}\mathrm{dx}d_{S}^{(1)}(x)e^{-ixt}\right\vert ^{2}\text{ ,}\\
A_{2}(t)  &  =\left\vert a_{2}(t)\right\vert ^{2}=\left\vert \int_{-\infty
}^{\infty}\mathrm{dx}d_{S}^{(2)}(x)e^{-ixt}\right\vert ^{2}\text{ ,}\\
A_{mix}(t)  &  =\frac{a_{1}^{\ast}(t)a_{2}(t)+a_{1}(t)a_{2}^{\ast}(t)}%
{2}\text{ }=\operatorname{Re}\left[  a_{1}(t)a_{2}^{\ast}(t)\right]  \text{ ,}%
\end{align}
have been introduced.

Therefore, the decay probability density $h(t)=-p^{\prime}(t)$ can be
rewritten as:%
\begin{equation}
h(t)=-A_{1}^{\prime}(t)-2A_{mix}^{\prime}(t)-A_{2}^{\prime}(t)\text{ .}
\label{hnew}%
\end{equation}
This form is the starting point for the postulate about $h_{1}(t)$ and
$h_{2}(t)$ as presented in the next subsection.%
\begin{figure}
[ptb]
\begin{center}
\includegraphics[
height=2.8911in,
width=4.6484in
]%
{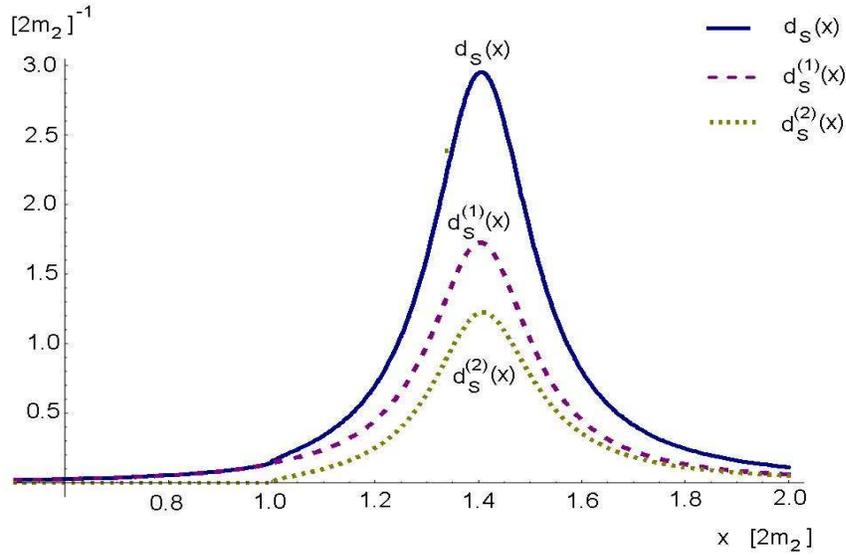}%
\caption{The partial distributions $d_{S}^{(1)}(x)$ and $d_{S}^{(2)}(x)$ are
plotted and compared to $d_{S}(x).$ The numerical values of Eq. (\ref{numval})
are used. }%
\end{center}
\end{figure}

\subsection{The decay probabilities $w_{1}(t)$ and $w_{2}(t)$ and the
densities $h_{1}(t)$ and $h_{2}(t)$ as functions of $a_{1}(t)$ and $a_{2}(t)$:
a conjectured solution and its properties}

\subsubsection{The postulate about $h_{i}(t)$ and $w_{i}(t)$}

The key functions in the present study are the densities of decay probability
per each channel, denoted as $h_{1}(t)$ and $h_{2}(t)$, which have been
introduced in the formal expressions (\ref{h1form}) and (\ref{h2form}).

We \emph{postulate} that the functions $h_{1}(t)$ and $h_{2}(t)$ can be
calculated as:%
\begin{equation}
h_{1}(t)=-A_{1}^{\prime}(t)-A_{mix}^{\prime}(t)\text{ ,} \label{h1}%
\end{equation}%
\begin{equation}
h_{2}(t)=-A_{2}^{\prime}(t)-A_{mix}^{\prime}(t)\text{ .} \label{h2}%
\end{equation}
The direct calculation of Eqs. (\ref{h1form}) and (\ref{h2form}) is a
complicated task in QFT, see also next Section where the explicit evaluation
in terms of Lee Hamiltonian(s) is performed. Here we argue that it is possible
to express the final result by using the introduced partial amplitudes
$a_{1}(t)$ and $a_{2}(t).$ Intuitively, we have split Eq. (\ref{hnew}) into
two pieces by assigning the mixing term $-2A_{mix}^{\prime}(t)$ in equal parts
to each channels. In the following subsections we show that Eqs.
(\ref{h1})-(\ref{h2}) fulfill all the required properties.

The decay probabilities $w_{1}(t)$ and $w_{2}(t)$, formally defined in Eqs.
(\ref{w1}) and (\ref{w2}), can be easily obtained from Eqs. (\ref{h1}%
)-(\ref{h2}) as:%
\begin{align}
w_{1}(t)  &  =A_{1}(0)+A_{mix}(0)-A_{1}(t)-A_{mix}(t)\text{ ,}\label{w1pr}\\
w_{2}(t)  &  =A_{2}(0)+A_{mix}(0)-A_{2}(t)-A_{mix}(t)\text{ .} \label{w2pr}%
\end{align}

In Fig. 7 the functions $h_{1}(t)$ and $h_{2}(t)$ as calculated from Eqs.
(\ref{h1})-(\ref{h2}) are plotted for the usual numerical example of Eq.
(\ref{numval}) and are compared to $h(t)$. Both functions vanish for $t=0$ and
have a qualitative similar behavior to the sum $h(t),$ reaching a maximum for
short times and then decreasing later on. It is also interesting to notice
that both functions show an oscillating behavior on top of the exponential
law.
\begin{figure}
[ptb]
\begin{center}
\includegraphics[
height=3.096in,
width=4.5005in
]%
{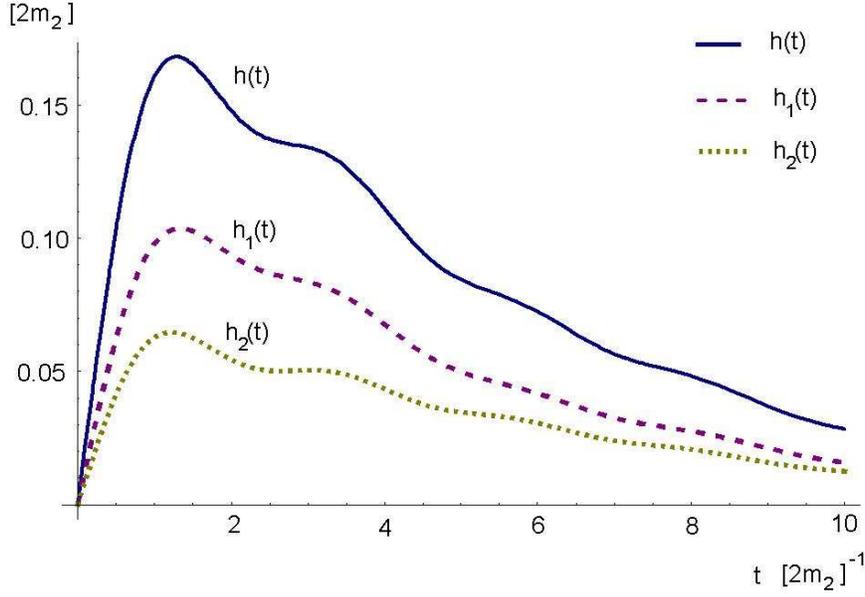}%
\caption{The partial decay probability densities $h_{1}(t)$ and $h_{2}(t)$ are
plotted and compared to $h(t)$. The numerical values of Eq. (\ref{numval}) are
used.}%
\end{center}
\end{figure}

\subsubsection{Sum}

The sum $h_{1}(t)+h_{2}(t)$ must obviously give $h(t)$:%
\begin{equation}
h(t)=h_{1}(t)+h_{2}(t)\text{ .} \label{sum}%
\end{equation}
In fact, the quantity $h(t)dt$ represents the probability that the particle
$S$ decays between $t$ and $t+dt,$ which must be the sum of the probability of
decaying into the first channel and into the second channel. This property,
which indeed has been our starting point leading to our postulate, is easily
verified, see Eq. (\ref{hnew}) and Eqs. (\ref{h1})-(\ref{h2}).

\subsubsection{Limits $g_{2}\rightarrow0$ and $g_{1}\rightarrow0$}

It is important to check that $h_{1}(t)$ and $h_{2}(t)$ fulfill the correct
limits when one of the coupling constant is sent to zero. In the limit
$g_{2}\rightarrow0$ one has, as expected, that%
\begin{equation}
h_{1}(t)=-A_{1}^{\prime}(t)\text{ },h_{2}(t)=0\text{ .}%
\end{equation}
Thus, only the first channel survives and $h(t)=h_{1}(t).$ Similarly, in the
limit $g_{1}\rightarrow0$ one has the specular situation%
\begin{equation}
h_{2}(t)=-A_{2}^{\prime}(t)\text{ , }h_{1}(t)=0\text{ .}%
\end{equation}

\subsubsection{Integrals and partial decay rates}

The functions $h_{1}(t)$ and $h_{2}(t)$ must fulfill the following
requirements:%
\begin{equation}
w_{1}(\infty)=\int_{0}^{\infty}\mathrm{dt}h_{1}(t)=\int_{0}^{\infty
}\mathrm{dx}d_{S}^{(1)}(x)\text{ ,} \label{h1int}%
\end{equation}%
\begin{equation}
w_{2}(\infty)=\int_{0}^{\infty}\mathrm{dt}h_{2}(t)=\int_{0}^{\infty
}\mathrm{dx}d_{S}^{(2)}(x)\text{ .} \label{h2int}%
\end{equation}
In fact, $w_{1}(\infty)=\int_{0}^{\infty}\mathrm{dt}h_{1}(t)$ represents the
probability that the particle decays in the first channel between $\left(
0,\infty\right)  $, a quantity which is also given by $\int_{0}^{\infty
}\mathrm{dx}d_{S}^{(1)}(x)$. (This is a typical result of the study of line
shapes, e.g. Ref. \cite{lupoder} and refs. therein.)

In order to prove Eq. (\ref{h1int}) we calculate%
\begin{equation}
\int_{0}^{\infty}\mathrm{dt}h_{1}(t)=\int_{0}^{\infty}\mathrm{dt}\left(
-A_{1}^{\prime}(t)-A_{mix}^{\prime}(t)\right)  =A_{1}(0)+A_{mix}(0).
\end{equation}
Taking into account that%
\begin{equation}
A_{1}(0)=\left(  \int_{-\infty}^{\infty}\mathrm{dx}d_{S}^{(1)}(x)\right)
^{2}\text{ ,}%
\end{equation}%
\begin{equation}
A_{mix}(0)=\left(  \int_{-\infty}^{\infty}\mathrm{dx}d_{S}^{(1)}(x)\right)
\left(  \int_{-\infty}^{\infty}\mathrm{dx}d_{S}^{(2)}(x)\right)  \text{ ,}%
\end{equation}
we obtain:%
\begin{equation}
A_{1}(0)+A_{mix}(0)=\left(  \int_{-\infty}^{\infty}\mathrm{dx}d_{S}%
^{(1)}(x)\right)  \left(  \int_{-\infty}^{\infty}\mathrm{dx}d_{S}%
^{(1)}(x)+\int_{-\infty}^{\infty}\mathrm{dx}d_{S}^{(2)}(x)\right)
=\int_{-\infty}^{\infty}\mathrm{dx}d_{S}^{(1)}(x)\text{ .}%
\end{equation}
In this way Eq. (\ref{h1int}) is proven. The validity of Eq. (\ref{h2int}) can
be easily proven following the same steps.

\subsubsection{The Breit-Wigner limit}

Also in the two-channel case it is important to study the Breit-Wigner limit.
The (non-relativistic) BW distributions $d_{S\text{-BW}}^{(1)}(x)$ and
$d_{S\text{-BW}}^{(2)}(x)$ of Eqs. (\ref{ds1})-(\ref{ds2}) are obtained just
as in Sec. 2.3:
\begin{align}
d_{S}^{(1)}(x)  &  \rightarrow d_{S\text{-BW}}^{(1)}(x)=\frac{\Gamma_{1}}%
{2\pi}\frac{1}{\left(  x-M\right)  ^{2}+\Gamma^{2}/4}\text{ ,}\\
d_{S}^{(2)}(x)  &  \rightarrow d_{S\text{-BW}}^{(2)}(x)=\frac{\Gamma_{2}}%
{2\pi}\frac{1}{\left(  x-M\right)  ^{2}+\Gamma^{2}/4}\text{ .}%
\end{align}
Obviously, $d_{S\text{-BW}}(x)=d_{S\text{-BW}}^{(1)}(x)+d_{S\text{-BW}}%
^{(2)}(x)$. The amplitudes $a_{1}(t)$ and $a_{2}(t)$ read in the BW limit%
\begin{equation}
a_{1}(t)\rightarrow a_{1,\text{BW}}(t)=\int_{-\infty}^{\infty}\mathrm{dx}%
d_{S}^{(1)}(x)e^{-ixt}=\frac{\Gamma_{1}}{\Gamma}e^{-\Gamma t/2}e^{iMt}\text{
,}%
\end{equation}%
\begin{equation}
a_{2}(t)\rightarrow a_{2,\text{BW}}(t)=\int_{-\infty}^{\infty}\mathrm{dx}%
d_{S}^{(2)}(x)e^{-ixt}=\frac{\Gamma_{2}}{\Gamma}e^{-\Gamma t/2}e^{iMt}\text{
.}%
\end{equation}
The quantity $A_{1},$ $A_{2}$ and $A_{\text{mix }}$ take the form
\begin{align}
A_{1}(t)  &  \rightarrow A_{1,\text{BW}}(t)=\left\vert a_{1,\text{BW}%
}(t)\right\vert ^{2}=\left(  \frac{\Gamma_{1}}{\Gamma}\right)  ^{2}e^{-\Gamma
t}\text{ ,}\\
A_{2}(t)  &  \rightarrow A_{2,\text{BW}}(t)=\left\vert a_{2,\text{BW}%
}(t)\right\vert ^{2}=\left(  \frac{\Gamma_{2}}{\Gamma}\right)  ^{2}e^{-\Gamma
t}\text{ ,}%
\end{align}%
\begin{equation}
A_{mix}(t)\rightarrow A_{mix\text{,BW}}(t)=\frac{a_{1,\text{BW}}^{\ast
}(t)a_{2,\text{BW}}(t)+a_{1,\text{BW}}(t)a_{2\text{,BW}}^{\ast}(t)}{2}%
=\frac{\Gamma_{1}\Gamma_{2}}{\Gamma^{2}}e^{-\Gamma t}.
\end{equation}
Out of these equations we can easily evaluate the functions $h_{1}$ and
$h_{2}$ in the BW limit:%
\begin{align}
h_{1}  &  \rightarrow h_{1,\text{BW}}(t)=\left(  \frac{\Gamma_{1}^{2}}{\Gamma
}+\frac{\Gamma_{1}\Gamma_{2}}{\Gamma}\right)  e^{-\Gamma t}=\Gamma
_{1}e^{-\Gamma t}\text{ ,}\\
h_{2}  &  \rightarrow h_{2,\text{BW}}(t)=\left(  \frac{\Gamma_{2}^{2}}{\Gamma
}+\frac{\Gamma_{1}\Gamma_{2}}{\Gamma}\right)  e^{-\Gamma t}=\Gamma
_{2}e^{-\Gamma t}\text{ .}%
\end{align}
The decay probabilities $w_{1}(t)$ and $w_{2}(t)$ take the form:%

\begin{align}
w_{1}(t)  &  \rightarrow w_{1,\text{BW}}(t)=\frac{\Gamma_{1}}{\Gamma}\left(
1-e^{-\Gamma t}\right)  \text{ ,}\\
w_{2}(t)  &  \rightarrow w_{2,\text{BW}}(t)=\frac{\Gamma_{2}}{\Gamma}\left(
1-e^{-\Gamma t}\right)  \text{ .}%
\end{align}
In particular, when calculating the decay fractions in a given channel, we
obtain the expected results%
\begin{align}
w_{1,\text{BW}}(\infty)  &  =\int_{0}^{\infty}\mathrm{dt}h_{1,\text{BW}%
}(t)=\frac{\Gamma_{1}}{\Gamma}\text{ ,}\\
w_{2,\text{BW}}(\infty)  &  =\int_{0}^{\infty}\mathrm{dt}h_{2,\text{BW}%
}(t)=\frac{\Gamma_{2}}{\Gamma}\text{ .}%
\end{align}

In Fig. 8 the functions $h_{1}(t)$ and $h_{2}(t)$ of Eqs. (\ref{h1}%
)-(\ref{h2}) are plotted separately and compared to the corresponding
exponential functions $h_{1,\text{BW}}(t)$ and $h_{2,\text{BW}}(t)$. It is
visible that also in each channel the deviations from the BW limit are not
negligible for small times, but become smaller and smaller with increasing
$t.$ It is also interesting to notice that the two functions $h_{1}(t)$ and
$h_{2}(t)$ oscillate around the corresponding BW limit in a different and
peculiar way.%

\begin{figure}
[ptb]
\begin{center}
\includegraphics[
height=2.3947in,
width=6.2249in
]%
{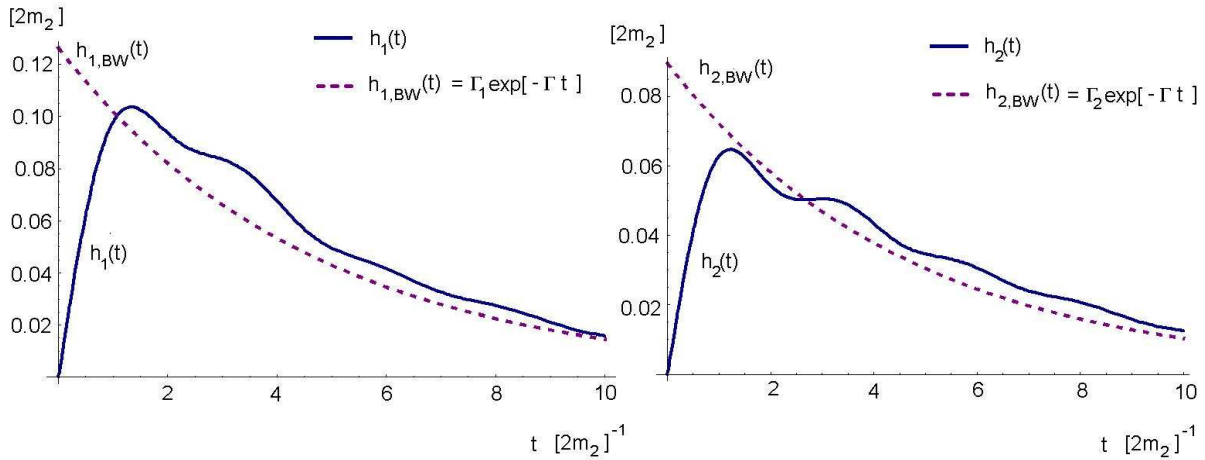}%
\caption{Left: the decay probability density into the first channel $h_{1}(t)$
is plotted together with the BW limit $h_{1,\text{BW}}(t)=\Gamma_{1}e^{-\Gamma
t}.$ Right: the same for $h_{1}(t)$ and $h_{2,\text{BW}}(t)$. In both cases
the numerical values of Eq. (\ref{numval}) have been used.}%
\end{center}
\end{figure}

The results obtained in the BW limit are expected: the usual exponential law
is recovered and the different channels are proportional to the corresponding
partial decay width $\Gamma_{i}$. In the BW limit the ratio $R(t)$ defined in
Eq. (\ref{r}) becomes a constant:%
\begin{equation}
R(t)=\frac{h_{1}(t)}{h_{2}(t)}\rightarrow R_{\text{BW}}(t)=\frac{\Gamma_{1}%
}{\Gamma_{2}}=const\text{ .}%
\end{equation}
This result does not hold true in the general case, in which $R(t)$ varies
with $t$. In Fig. 9 the function $R(t)$ and the constant BW limit
$R_{\text{BW}}(t)=\Gamma_{1}/\Gamma_{2}$ are shown: the deviation of the exact
result $R(t)$ from the BW limit is sizable. There are time intervals in which
the production of $\varphi_{1}\varphi_{1}$ pairs is enhanced with respect to
$\varphi_{2}\varphi_{2}$ pairs, and there are time intervals in which the
opposite is true. This is an extremely interesting behavior which shall be
studied in detail in the future. The reason for the large departure from the
constant BW result can be also traced back to the oscillating behavior of both
functions $h_{1}(t)$ and $h_{2}(t).$ The frequencies of the oscillations are
different, and thus the ratio $R(t)$ results as a complicated outcome.
Moreover, the deviations are still large also for large times. This is due to
the fact that for $t\gg\tau_{\text{BW}}$ the quantities $h_{1}(t)$ and
$h_{2}(t)$ are both small numbers: even small variations from the exponential
limit cause large deviations of the ratio $R(t)$ from the constant value
$\Gamma_{1}/\Gamma_{2}$. The full study of the large time behavior of the
function $R(t)$ represents also an interesting outlook for future studies.%

\begin{figure}
[ptb]
\begin{center}
\includegraphics[
height=2.8556in,
width=5.5677in
]%
{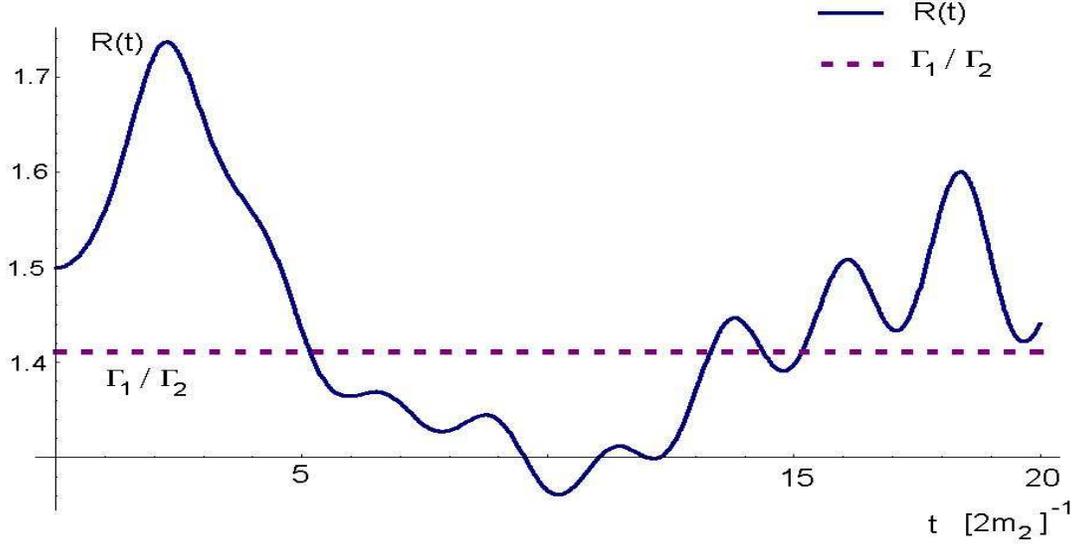}%
\caption{The ratio $R(t)=h_{1}(t)/h_{2}(t)$ (solid line) is plotted and
compared to the constant BW limit $R_{\text{BW}}(t)=\Gamma_{1}/\Gamma_{2}$
(dashed line) Large deviations, which persist also at large times, take place.
As usual, the numerical values (\ref{numval}) have been used.}%
\end{center}
\end{figure}

Finally, in Fig. 10 we plot the ratios
\[
\frac{h_{i}(t)}{h(t)}\text{ and}\frac{w_{i}(t)}{w(t)}\text{ , }i=1,2\text{ .}%
\]
Both functions tend to the constant value $\Gamma_{i}/\Gamma$ in the BW limit.
One can see that $h_{i}(t)/h(t)$ oscillate and that $w_{i}(t)/w(t)$ tend to a
constant limit which are not coincident with the naive BW expectations, thus
showing that deviations persist also for large times.%

\begin{figure}
[ptb]
\begin{center}
\includegraphics[
height=2.6394in,
width=5.9672in
]%
{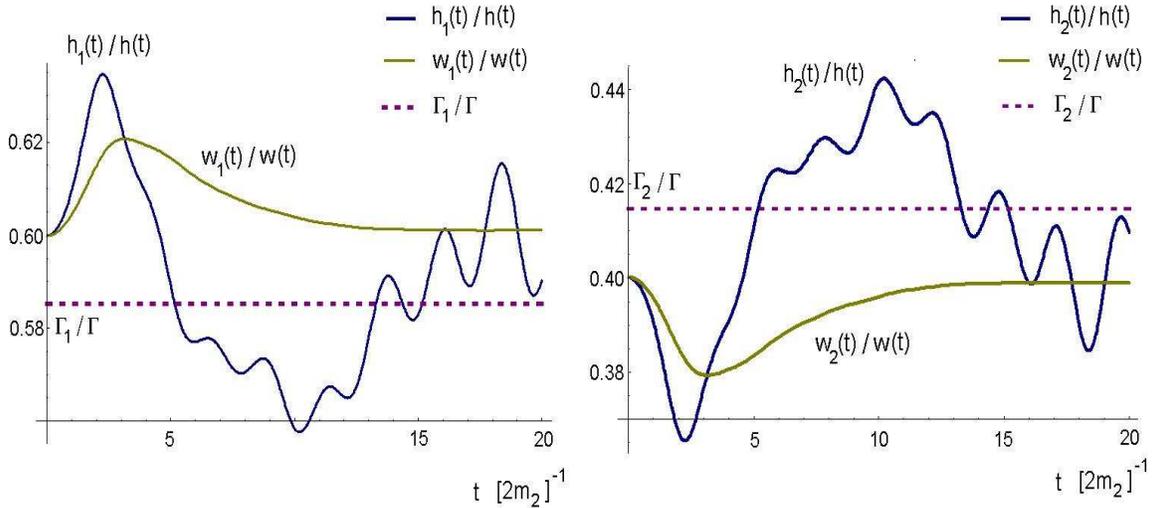}%
\caption{Left: $h_{1}(t)/h(t)$ and $w_{1}(t)/w(t)$ have been plotted and
compared to the constant BW limit $\Gamma_{1}/\Gamma$. Right: $h_{2}(t)/h(t)$
and $w_{2}(t)/w(t)$ have been plotted and compared to the BW limit $\Gamma
_{2}/\Gamma$. The numerical values of Eq. (\ref{numval}) have been used.}%
\end{center}
\end{figure}

\subsubsection{Invariance under time-reversal}

The underlying theory described by the Lagrangian of Eq. (\ref{lag}) is
invariant under time reversal; therefore, the physical quantities under study
should also reflect this basic symmetry. The amplitude $a(t)$ defined in Eq.
(\ref{a(t)}) fulfills the equation $a(-t)=a^{\ast}(t)$. As a consequence, the
survival probability $p(t)=a^{\ast}(t)a(t)$ is invariant under time reversal:
$p(-t)=p(t)$.

Also the decay probability $w(t)$ in Eq. (\ref{intothsv}) is invariant under
time reversal for the very same reason: $w(-t)=w(t)$. Notice that the decay
probability density $h(t)$ is odd under time reversal: $h(-t)=-h(t).$ In this
way the invariance of the physical quantity $h(t)dt,$ and therefore $w(t)$, is assured.

The invariance under time reversal is satisfied also by the decay
probabilities $w_{1}(t)$ and $w_{2}(t)$, as the formal expressions in Eqs.
(\ref{w1}) and (\ref{w2}) show: $w_{1}(-t)=w_{1}(t),$ $w_{2}(-t)=w_{2}(t)$.
Due to the mathematical property $a_{1,2}(-t)=a_{1,2}^{\ast}(t)$ also the
expressions for $w_{1}(t)$ and $w_{2}(t)$ in Eqs. (\ref{w1pr}) and
(\ref{w2pr}) fulfill the requirement of invariance under time reversal. Just
as for the function $h(t)$, the decay probability densities $h_{1}(t)$ and
$h_{2}(t)$ must be odd under time reversal: $h_{1}(-t)=-h_{1}(t)$ and
$h_{2}(-t)=-h_{2}(t).$ This is easily verified for Eqs. (\ref{h1}) and
(\ref{h2}).

\subsubsection{On the uniqueness of the postulated solution}

Eqs. (\ref{h1})-(\ref{h2}) fulfill all the required conditions to represent
the decay probability densities in the first and the second channel. However,
it is important to study the uniqueness of presented solutions.

A different combination $\tilde{h}_{1}(t)=-A_{1}^{\prime}(t)-\alpha
A_{mix}^{\prime}(t)$, $\tilde{h}_{2}(t)=-A_{2}^{\prime}(t)-\beta
A_{mix}^{\prime}(t)$ with $\alpha+\beta=2$ and $\alpha,\beta\neq1$ would
obviously satisfy the constraint given by the sum (\ref{sum}) (Sec. 4.3.2) and
the limits $g_{1},g_{2}\rightarrow0$ (Sec. 4.3.3), but would not fulfill the
constraints given by the integrals (\ref{h1int})-(\ref{h2int}) and would not
reproduce the correct BW limit. Therefore, only the case $\alpha=\beta=1$ is allowed.

More in general, one can try to find alternative solutions by writing down the
functions%
\begin{align}
\tilde{h}_{1}(t)  &  =h_{1}(t)+\delta(t)\\
\tilde{h}_{2}(t)  &  =h_{2}(t)-\delta(t)
\end{align}
where the `hypothetical' function $\delta(t)\equiv\delta(g_{1},m_{1}%
,g_{2},m_{2},t)$ has been introduced. While the sum constraint of Eq.
(\ref{sum}) is fulfilled by construction, the function $\delta(g_{1}%
,m_{1},g_{2},m_{2},t)$ must vanish for $g_{1}\rightarrow0$ and for
$g_{2}\rightarrow0$ (Sec. 4.3.3) and the integral over time must vanish,
$\int_{0}^{\infty}\delta(t)dt=0$, in order that Eqs. (\ref{h1int}%
)-(\ref{h2int}) still hold. Then, $\delta(t)$ must also vanish in the BW limit
to be in agreement with Sec. 4.3.5. Moreover, being $\tilde{h}_{1}(t)$ and
$\tilde{h}_{2}(t)$ positive definite, $\left\vert \delta(t)\right\vert \leq
h_{2}(t)$ when $\delta(t)>0$ and $\left\vert \delta(t)\right\vert \leq
h_{1}(t)$ when $\delta(t)<0.$ All these properties are automatically fulfilled
by the trivial solution $\delta(t)=0$, but the question is if other
possibilities exist.

A further property that $\delta$ must fulfill is obtained by studying the
exchange of the two particles. This operation amounts to exchange
$g_{1}\longleftrightarrow g_{2}$ and $m_{1}\longleftrightarrow m_{2}$. While
in this case $h_{1}\leftrightarrow h_{2}$, it must hold that:%
\begin{equation}
\delta(g_{1},m_{1},g_{2},m_{2},t)=-\delta(g_{2},m_{2},g_{1},m_{1},t)\text{ .}
\label{exchange}%
\end{equation}
If we consider, for instance, the function%
\begin{equation}
\delta(t)=\delta_{0}\frac{d}{dt}\operatorname{Im}\left[  a_{1}(t)a_{2}^{\ast
}(t)\right]  \label{deltaex}%
\end{equation}
where the constant $\delta_{0}$ is small enough to guarantee the semi-positive
definiteness of $\tilde{h}_{1}(t)$ and $\tilde{h}_{2}(t)$, it is easy to prove
that Eq. (\ref{deltaex}) would actually fulfill all the mentioned properties
from Sec. 4.3.2 to Sec. 4.3.5 together with the exchange property of Eq.
(\ref{exchange}). However, it violates time-reversal, because it is even under
switch of $t$: $\delta(-t)=\delta(t)$ (while, as shown in Sec. 4.3.6, the
decay probability densities must be odd under time reversal). For this reason
the only possibility is $\delta_{0}=0$, i.e. the function $\delta(t)$ vanishes.

This discussion can be generalized. Writing $\delta(t)=\frac{d}{dt}\Delta(t)$
and making the additional assumption that the function $\Delta(t)$ can be
expressed as $\Delta(t)=\sum_{i,j}c_{ij}a_{i}(t)a_{j}^{\ast}(t)$ where the
coefficients $c_{ij}$ are pure numbers, it is not possible to build
$\delta(t)$ such that it is odd under time reversal and at the same time
fulfills Eq. (\ref{exchange}). This chain of arguments represent a partial
proof of the validity of Eqs. (\ref{h1})-(\ref{h2}).

Summarizing, all these arguments enforce our point of view that the Eqs.
(\ref{h1})-(\ref{h2}), and therefore Eqs. (\ref{w1pr})-(\ref{w2pr}), are the
correct answer to the problem. However, the final answer can only be given if
one can mathematically prove the validity of the equations (\ref{h1}%
)-(\ref{h2}) under general assumptions. Presently, we did not achieve this
point, which remains an important task for future studies. However, by using
the formalism of the Lee Hamiltonians it is possible to find the exact
expressions for $h_{i}(t)$ and $w_{i}(t),$ which allow for a numerical
verification of Eqs. (\ref{h1})-(\ref{h2}), see Sec. 5.

\subsubsection{Generalizations}

The present discussion can be easily generalized to the case of a $n$-channel
decay: this case is discussed in Appendix B. Moreover, the formalism is not
restricted to scalar field theories only but can be easily generalized to each
QFT Lagrangian.

\subsubsection{Deviations from the exponential decay law and measurements}

In this work we focus our attention to the undisturbed decay probabilities of
the system: after the creation at $t=0$ the unstable system $\left\vert
S\right\rangle $ evolves according to the Hamiltonian of the system which
control its decay. We do not attempt here to include the measurement into the
mathematical treatment. This is surely an interesting subject which is left as
an outlook.

This problem has been first investigated in Refs.
\cite{ghirardi,degasperis,fonda}. In particular, in\ Ref. \cite{degasperis} it
has been pointed out that random measurements occurring with a certain mean
frequency $\nu$ make the decay still appear to be an exponential, even if the
frequency is large enough to probe the non-exponential decay time. This is
indeed a generalization of the simple discussion of Sec.2.5, where the
quantum-Zeno effect has been presented as an outcome of pulsed measurements.
It is assumed that the measurement lets the state collapse back into the
unstable state $\left\vert S\right\rangle $, which also implies a reset of the
clock to the initial value of $t=0.$ In this framework, the deviations from
the exponential decay law studied in this work would manifest in an
exponential decay laws with a ratio which can be different from $\Gamma
_{1}/\Gamma_{2}$ in dependence of the (mean) frequency of the observations.

On the other hand, future studies in this direction should also take into
account the theory of measurement, see Ref. \cite{koshino} and refs. therein,
in which it is shown that the influence of the measurement apparatus and its
response time can have a non-negligible effect. As studied also in\ Refs.
\cite{hotta,delgado}, the outcome may also depend on the details of the
experimental apparatus. A detailed investigation of the `collapse' beyond the
simple projection postulate is also an interesting direction of study in
connection with non-exponential decay laws. Another interesting line of
research is to investigate in more details the problem of sequential decay
processes analyzed in Ref. \cite{fonda} by using the approach described in the
present manuscript.

\section{Two-channel case: QM and link to QFT}

\subsection{Exact determination of the decay probabilities $w_{1}(t)$ and
$w_{2}(t)$ in QM by using the nonrelativistic LH}

We now study the two-channel decay problem in the QM framework by using the
Lee Hamiltonian $H$ introduced in Eq. (\ref{hleenr}).

Due to the formal analogy of the QM-LH and the QFT approaches, we can repeat
here all the steps done in Sec. 4.2 and 4.3 leading to the postulated
solutions for $w_{i}(t)$ and $h_{i}(t)$. They read:
\begin{align}
w_{i}(t)  &  =A_{i}(0)+A_{mix}(0)-A_{i}(t)-A_{mix}(t)\text{ , }h_{i}%
(t)=-w_{i}^{\prime}(t)\text{ ,}\nonumber\\
A_{i}(t)  &  =\left\vert a_{i}(t)\right\vert ^{2}\text{ , }A_{mix}%
(t)=\operatorname{Re}\left[  a_{1}(t)a_{2}^{\ast}(t)\right]  \text{ ,}
\label{wilee}%
\end{align}
where
\begin{equation}
a_{i}(t)=\left\vert \int_{-\infty}^{\infty}\mathrm{dx}d_{S}^{(i)}%
(x)e^{-ixt}\right\vert ^{2}\text{ , }d_{S}^{(i)}(x)=\frac{1}{\pi}\left\vert
\frac{g_{i}^{2}\operatorname{Im}\left[  \Sigma_{i}(x)\right]  }{(x-M_{0}%
+\operatorname{Re}[\Pi(x)])^{2}+\operatorname{Im}^{2}\left[  \Pi(x)\right]
}\right\vert \text{ .} \label{wilee2}%
\end{equation}
(See Sec. 3.1 for the definitions of the functions $\Sigma_{i}(x)$ and
$\Pi(x)$ in the nonrelativistic QM-LH case).

In the present QM-LH treatment we can calculate \emph{exactly} the functions
$w_{i}(t)$. To this end we first write the formal expression for the decay
probability $w_{i}(t)$ in the $i$-th channel:%

\begin{equation}
w_{i}(t)=\int_{0}^{t}h_{i}(u)du=\sum_{\mathbf{k}}\left\vert \left\langle
i,\mathbf{k}\left\vert e^{-iHt}\right\vert S\right\rangle \right\vert
^{2}\text{ , }i=1,2. \label{wileenr}%
\end{equation}
We thus need to evaluate the matrix element
\begin{equation}
\left\langle i,\mathbf{k}\left\vert e^{-iHt}\right\vert S\right\rangle \text{
.}%
\end{equation}
Proceeding as in Sec. 3.1 we obtain the following expression:%
\begin{equation}
\left\langle i,\mathbf{k}\left\vert e^{-iHt}\right\vert S\right\rangle \text{
}=\frac{i}{2\pi}g_{i}\frac{f_{i}(\mathbf{k)}}{\sqrt{V}}\int_{-\infty}^{\infty
}dE\frac{G_{S}(E)}{E-\omega_{i}(\mathbf{k})+i\varepsilon}\text{ .}%
\end{equation}
Introducing the dispersion relation of $G_{S}(E)$, see Eq. (\ref{disprel}),
and performing the subsequent integral, we arrive at the following equation:
\begin{equation}
\left\langle i,\mathbf{k}\left\vert e^{-iHt}\right\vert S\right\rangle
=g_{i}\frac{f_{i}(\mathbf{k)}}{\sqrt{V}}\int_{-\infty}^{\infty}dxd_{S}%
(x)\frac{e^{-i\omega_{i}(\mathbf{k})t}-e^{-ixt}}{\omega_{i}(\mathbf{k}%
)-x}\text{ .}%
\end{equation}
Using now Eq. (\ref{wileenr}) we obtain the final, exact expression for
$w_{i}(t)$ (in the continuos limit) as:%
\begin{equation}
w_{i}^{\text{exact}}(t)=\int\frac{d^{3}k}{(2\pi)^{3}}g_{i}^{2}f_{i}%
^{2}(\mathbf{k)}\left\vert \int_{-\infty}^{\infty}dxd_{S}(x)\frac
{e^{-i\omega_{i}(\mathbf{k})t}-e^{-ixt}}{\omega_{i}(\mathbf{k})-x}\right\vert
^{2}\text{ , }h_{i}^{\text{exact}}(t)=-\frac{dw_{i}^{\text{exact}}(t)}%
{dt}\text{ .} \label{wiex}%
\end{equation}

We have now compared the expressions in Eq. (\ref{wiex}) and Eqs.
(\ref{wilee})-(\ref{wilee2}) for different parametrizations of the functions
$\omega_{i}(\mathbf{k})$ and $f_{i}(\mathbf{k})$. In particular, we have used:

(i) A generalized version of the two-pole model developed in Ref.
\cite{facchiprl}, in which two open channels are included. The numerical
values have been varied in order to check dependence on them.

(ii) Modifications of the two-pole model including energy thresholds.

(iii) Different choices for the loop functions $f_{i}(\mathbf{k})$.

In all these cases we have verified that the two expressions (\ref{wiex}) and
(\ref{wilee})-(\ref{wilee2}) are numerically indistinguishable. All the
results obtained in the present QM-LH approach lead us to conclude that the
postulated expressions in\ Eqs. (\ref{wilee})-(\ref{wilee2}) (which have the
same form as the QFT-like expressions in Sec. 4.3) represent --at least-- a
very good approximation of the exact results. A proof of the complete
equivalence (or eventually of the -albeit small- deviations) of these
expressions is left for the future. In particular, one should check the
validity of the Ansatz in\ Eqs. (\ref{wilee})-(\ref{wilee2}) under more
complicated circumstances, involving a numerical check over a large range of
parameters and with different forms of the functions $f_{i}(\mathbf{k})$.

In the end, it is also important to stress that the qualitative behavior of
the functions $h_{i}(t)$ and $R(t)$ as evaluated in the present QM-LH approach
is very similar to the plots presented in this manuscript: $h_{i}(t)$ vanish
for $t\rightarrow0$ and then tend to the BW limit for large times. Typical
oscillations are present and the ratio $R(t)$ is varying abruptly as in Fig. 9.

\subsection{Generalization to the relativistic case and link to QFT}

We now study the two-channel problem using the formalism developed in Sec.
3.2, in which a relativistic generalization of the QM-LH approach was developed.

The solutions postulated in Sec. 4.2 and 4.3 takes the following form in the
present case:%

\begin{align}
w_{i}(t)  &  =A_{i}(0)+A_{mix}(0)-A_{i}(t)-A_{mix}(t)\text{ , }h_{i}%
(t)=-w_{i}^{\prime}(t)\text{ ,}\nonumber\\
A_{i}(t)  &  =\left\vert a_{i}(t)\right\vert ^{2}\text{ , }A_{mix}%
(t)=\operatorname{Re}\left[  a_{1}(t)a_{2}^{\ast}(t)\right]  \text{ ,}
\label{wileerel}%
\end{align}
where
\begin{equation}
a_{i}(t)=\left\vert \int_{-\infty}^{\infty}\mathrm{dx}d_{S}^{(i)}%
(x)e^{-ixt}\right\vert ^{2}\text{ , }d_{S}^{(i)}(x)=\frac{2x}{\pi}\left\vert
\frac{g_{i}^{2}\operatorname{Im}\left[  \Sigma_{i}(x)\right]  }{(x^{2}%
-M_{0}^{2}+\operatorname{Re}[\Pi(x)])^{2}+\operatorname{Im}^{2}\left[
\Pi(x)\right]  }\right\vert \text{ .} \label{wileerel2}%
\end{equation}
(See Sec. 3.2 for the definitions of the functions $\Sigma_{i}(x)$ and
$\Pi(x)$ in the relativistic LH treatment). Note that, when using the
identifications of Eqs. (\ref{eq1})-(\ref{eq2}), the previous equations
coincide with those in Sec. 4.3.

By following the same steps of Sec. 5.1 (that is, taking into account the
relativistic form of the propagators) we arrive at the following `exact'
expressions for $w_{i}(t)$ for the relativistic LH approach:%

\begin{equation}
w_{i}^{\text{exact}}(t)=\int\frac{d^{3}k}{(2\pi)^{3}}g_{i}^{2}f_{i}%
^{2}(\mathbf{k)}\left\vert \int_{-\infty}^{\infty}dxd_{S}(x)\frac
{e^{-i\omega_{i}(\mathbf{k})t}-e^{-ixt}}{\omega_{i}^{2}(\mathbf{k})-x^{2}%
}\right\vert ^{2}\text{ , }h_{i}^{\text{exact}}(t)=-\frac{dw_{i}%
^{\text{exact}}(t)}{dt}\text{ .} \label{wiexrel}%
\end{equation}
Also in this case we have numerically compared the conjectured solutions of
Eqs. (\ref{wileerel})-(\ref{wileerel2}) with the exact result in\ Eq.
(\ref{wiexrel}) and found them equal.

In particular, when using the identifications of Eqs. (\ref{eq1})-(\ref{eq2})
which lead to a mathematical equivalence of the present LH approach with the
QFT model described in Secs. 2 and 4, we have recalculated all the numerical
results presented in Figs. 2-10 with the help of the exact expressions in\ Eq.
(\ref{wiexrel}): as expected, no deviation was found. This fact reinforces our
point of view that the expressions represent the exact result (or a very
accurate approximation of it).

\section{Conclusions}

In this work we have studied the decay law(s) as function of time of an
unstable particle $S$, which can decay into two (or more) decay channels. We
have done this investigation in two different theoretical contexts: using a
relativistic quantum field theoretical Lagrangian and using a quantum
mechanical approach expressed in terms of a Lee Hamiltonian.

In the first part of the manuscript we have reviewed general properties of the
survival probability $p(t)$ and the decay probability density $h(t)=-p^{\prime
}(t)$ (Fig. 1-5, Sec. 2 and Sec. 3) for both the QFT and the QM cases: the
non-exponential behavior of the survival probability $p(t)$ for short times,
the Breit-Wigner limit leading to the usual exponential decay law, and the
quantum Zeno effect have been described. Formally, the QFT and the QM-LH
approaches are very similar, the main difference being the presence of
relativistic propagator in the former case and nonrelativistic propagators in
the latter one. To overcome this difference we have also developed a simple
relativistic generalization of the QM-LH approach in which the same equations
of the QFT case (can) arise.

In the second part of the manuscript (Secs. 4 and 5) we have concentrated our
attention on the peculiarities of the decay into each one of the two (or more)
available open channels. In the two-channel case the mathematical expressions
for the decay probability densities into each channel, $h_{1}(t)$ and
$h_{2}(t)$, and of their corresponding integrated quantities $w_{1}%
(t)=\int_{0}^{t}\mathrm{du}h_{1}(u)$ and $w_{2}(t)=\int_{0}^{t}\mathrm{du}%
h_{2}(u)$ have been discussed. Namely, $h_{i}(t)dt$ represents the probability
that the unstable particle $S$ decays into the $i$-th channel between $t$ and
$t+dt,$ while $w_{i}(t)$ represent the probability that the particle decays
into the $i$-th channel between $0$ and $t.$ Obviously, $w_{1}(t)+w_{2}%
(t)=1-p(t).$

In particular, in QFT (Sec. 4) we did not evaluate $h_{i}(t)$ and $w_{i}(t)$
exactly, but we have postulated a solution (Eqs. (\ref{h1})-(\ref{h2}) in Sec.
4.3, see also Figs. 7-8). We have shown that the proposed expressions for
$h_{i}(t)$ and $w_{i}(t)$ fulfill all the required properties, but we could
prove the mathematical exactness of the results only under a restricted set of
hypotheses, see Sec. 4.3.7 and below. We have also shown that the conjectured
solutions show strong deviations from the Breit-Wigner limit, in which the
decay law is exponential. An important quantity that we have studied is the
ratio of probability densities $R(t)=h_{1}(t)/h_{2}(t)$. While this quantity
is constant in the Breit-Wigner limit and equals the ratio of the tree-level
decay widths ($R(t)\rightarrow\Gamma_{1}/\Gamma_{2}$), it turns out that the
full expression is in general not constant and may deviate sizably from the
constant limit, see Fig. 9. The functions $h_{i}(t)/h(t)$ and $w_{i}(t)/w(t)$
show also the peculiar characteristics of the two-channel problem and have
been plotted in Fig. 10.

In view of the fact that the conjectured solutions in QFT could not be
analytically demonstrated in the general case, we have studied the same
problem in the formally very similar (up to the presence of nonrelativistic
propagators) approach of a nonrelativistic QM-LH approach. In this case,
besides the postulated solutions, it is also possible to derive the
$\emph{exact}$ form for the functions $h_{i}^{exact}(t)$ and $w_{i}%
^{exact}(t)$. A numerical analysis was performed in order to show that the
proposed solutions $h_{i}(t)$ and the exact ones $h_{i}^{exact}(t)$ coincide,
thus proving that the proposed solutions represent --at least-- a very good
approximation of the exact results. When studying the relativistic
generalization of the Lee Hamiltonian approach, in which the equations can be
chosen to coincide with the QFT study, we have once more numerically verified
that the expressions postulated in the QFT context are valid: using the very
same parameters of Eq. (\ref{numval}) we have checked the correctness of Figs. 2-10.

The are various possible outlooks for the future, which can be divided into
mathematical developments and physical applications.

On the mathematical side, one can perform the following studies: (i)
Mathematical proof of the correctness of the postulated expressions for
$h_{i}(t)$ and $w_{i}(t)$ presented in Sec. 4.3 in the QFT case. If the
proposed solutions are not exact, it would be necessary to investigate why
they offer such a good numerical approximation to the problem. In this
context, one needs to check the numerical validity of the postulated Ansatz
under more general conditions. (ii) Inclusion of higher orders in the
calculation of the self-energy diagrams of the QFT case, thus going beyond
Fig. 1.b. (iii) Extension to (strictly) renormalizable Lagrangians.

On a phenomenological level various investigations are possible: (i) Study of
physical processes in the framework of the Standard Model, where decays of
unstable particles into many channels have been observed and precisely
measured. To this end the mathematical development of the mathematical outlook
(iii) is necessary. (ii) Applications to weak decays (such as electronic
capture decays) of atoms. This analysis could be important to study in detail
the properties of oscillations such as the ones experimentally found in Ref.
\cite{litvinov}, see also Ref. \cite{gsiano} for a related theoretical
discussion. (iii) Applications to hadron physics: the deviations from the
non-exponential law are large in this case \cite{zeno1}, although they last
very short due to the short-living nature of (most) hadrons. However,
deviations from the exponential decay law might be interesting in connection
to an expanding gas of hadrons produced in heavy ion collisions
\cite{bleicher}. (iv) A peculiar situation takes place when one channel is
kinematically not allowed, $M<2m_{2}.$ When $M$ is just slightly smaller then
$2m_{2}$, the finite width of the unstable state $S$ due to its decay into the
first channel allows also the decay into the --nominally forbidden-- second
one. This situation is physically realized in the case of the scalar mesons
$a_{0}(980)$ and $f_{0}(980)$, see details in Refs.
\cite{lupo1,amslerrev,lupoder,last} and refs. therein. The study of the
temporal evolution in such a case represents an interesting outlook.

As a last remark, it is conceivable that the decay into different channels can
be experimentally studied with the procedure of Ref. \cite{raizen1};
asymmetric barriers should be used, in order to allow for different tunneling
probabilities is different directions. Then, it would be possible to test in a
experimentally controlled way quantities such as the ratio $R(t)=h_{1}%
(t)/h_{2}(t)$ described in this work. More in general, in connection with the
experiments, one should also go beyond the undisturbed time-evolution of the
system and include the measuring procedure as a part of the physical
description of the decay.

\bigskip

\textbf{Acknowledgments: }The author thanks Giuseppe Pagliara for cooperation
and many valuable discussions on the subject and Thomas Wolkanowski for useful remarks.

\bigskip

\bigskip\appendix

\section{The loop function}

In order to regularize the self-energy diagram of Fig. 1.b we introduce at
each vertex the vertex-function $\tilde{\phi}(q)$. At the level of the
Lagrangian formulation this can be achieved by rendering the Lagrangian
nonlocal, see details in Refs. \cite{lupo1,nl} and refs. therein. The integral
in Eq. (\ref{self}) takes the form%

\begin{equation}
\Sigma(p^{2},m^{2})=-i\int\frac{d^{4}q}{(2\pi)^{4}}\frac{\tilde{\phi}(q)^{2}%
}{\left[  \left(  \frac{p+2q}{2}\right)  ^{2}-m^{2}+i\varepsilon\right]
\left[  \left(  \frac{p-2q}{2}\right)  ^{2}-m^{2}+i\varepsilon\right]  }\text{
.} \label{intlupo}%
\end{equation}
A general property of $\Sigma(x,m^{2})$ follows from the optical theorem:
\[
(\sqrt{2}g)^{2}\mathrm{Im}[\Sigma(x,m^{2})]=x\Gamma^{\text{t-l}}(x,m,g)\left[
\tilde{\phi}(q=(0,\mathbf{q}))\right]  ^{2}\text{,}%
\]
where
\begin{equation}
\mathbf{q}^{2}=\frac{x^{2}}{4}-m^{2}\text{ .}%
\end{equation}
Due to spatial isotropy, the function $\tilde{\phi}(q=(0,\mathbf{q}))$ must
depend on $\mathbf{q}^{2}.$

For a generic $\tilde{\phi}(q)$, the tree-level decay width is indeed not
given by Eq. (\ref{tldf}) but takes the following modified form:
\begin{equation}
\Gamma^{\text{t-l}}(x,m,g)\rightarrow\Gamma^{\text{t-l}}(x,m,g)\left[
\tilde{\phi}(q=(0,\vec{q}))\right]  ^{2}\text{ .}%
\end{equation}

When assuming that $\tilde{\phi}(q)=\tilde{\phi}(\mathbf{q})$ (i.e. when we
drop the dependence on $q^{0}$), we can perform, in the rest frame of the $S$
particle ($p=(x,0)$), the integral over $q^{0}$ in Eq. (\ref{intlupo}),
obtaining:%
\begin{equation}
\Sigma(x^{2},m^{2})=\int\frac{d^{3}q}{(2\pi)^{3}}\frac{\tilde{\phi}%
^{2}(\mathbf{q})}{\sqrt{\mathbf{q}^{2}+m^{2}}\left(  4(\mathbf{q}^{2}%
+m^{2})-x^{2}-i\varepsilon\right)  }\text{ .}\label{luporeg}%
\end{equation}
In this work we make the following simple choice for numerical evaluations:
\begin{equation}
\tilde{\phi}(q)=\tilde{\phi}(\mathbf{q})=\theta(\Lambda^{2}-\mathbf{q}%
^{2})\text{ ,}%
\end{equation}
i.e. we work with a sharp cutoff $\Lambda.$ (In general, the use of smooth
functions affects the results only slightly, thus showing a weak dependence on
the adopted vertex function. An explicit calculation has been presented in
Ref. \cite{zenoproc}). In this case, by performing also the spatial
integration we obtain the explicit result:
\begin{equation}
\Sigma(x,m^{2})=\frac{-\sqrt{4m^{2}-x^{2}}}{8\pi^{2}x}\arctan\left(
\frac{\Lambda x}{\sqrt{\Lambda^{2}+m^{2}}\sqrt{4m^{2}-x^{2}}}\right)
-\frac{1}{8\pi^{2}}\log\left(  \frac{m}{\Lambda+\sqrt{\Lambda^{2}+m^{2}}%
}\right)  \text{ .}%
\end{equation}

Following comments are in order:

(i) The imaginary part of the self-energy amplitude $\operatorname{Im}%
[\Sigma(x^{2},m^{2})]$ is zero for $0<x<2m$ and nonzero starting at threshold.

(ii) The real part $(\sqrt{2}g)^{2}\mathrm{Re}[\Sigma(x^{2},m^{2})]$ is
nonzero below and above threshold and depends explicitly on the cutoff
$\Lambda$.

(iii) For the case $\tilde{\phi}(q)=\theta(\Lambda^{2}-\mathbf{q}^{2})$, the
optical theorem takes the form%
\begin{equation}
(\sqrt{2}g)^{2}\mathrm{Im}[\Sigma(x^{2},m^{2})]=x\Gamma^{\text{t-l}%
}(x,m,g)\theta\left(  \sqrt{\Lambda^{2}+m^{2}}-\frac{x}{2}\right)  \text{,}%
\end{equation}
Besides the additional $\theta$ function, it does not depend on the cutoff
$\Lambda$. Note that Eq. (\ref{impi}) in the text is strictly speaking valid
only for $x<2\sqrt{\Lambda^{2}+m_{1}^{2}}$. This condition is numerically met
in Fig. 2 and Fig. 6.

(iv) The present choice of $\tilde{\phi}(q)$ breaks Lorentz invariance. This
is here not a problem, because we always work in the rest frame of $S.$ It is
also not difficult to generalize $\tilde{\phi}(q)$ in order that it is Lorentz
invariant and delivers the same results of this paper when the rest frame of
$S$ is considered.

As a last step we turn to the issue of the bare and renormalized masses: in
this work we started with the bare mass $M_{0}$ entering in the Lagrangian of
Eq. (\ref{lag}) and then we derived the renormalized mass $M$ in Eq.
(\ref{eqm}), arising upon the inclusion and the resummation of the self-energy
diagram of Fig. 1.b. Alternatively, one could have added a counterterm in Eq.
(\ref{lag}):
\begin{equation}
\mathcal{L\rightarrow L-}\frac{1}{2}CS^{2}\text{ with }C=R(M_{0}).
\end{equation}
In this way the mass equation takes the form%
\[
M^{2}-M_{0}^{2}-C+R(M)=0\rightarrow M=M_{0}\text{ .}%
\]
Clearly, all the physical results of this work would be unaffected by this
alternative procedure.

\section{The $n$-channel case}

In the $n$-channel case the interaction Lagrangian takes the form
\begin{equation}
\mathcal{L}_{int}=\sum_{i=1}^{n}g_{i}S\varphi_{i}^{2}\text{ ,}%
\end{equation}
where the particle $\varphi_{i}$ has a mass $m_{i}$. The spectral function (or
mass distribution) $d_{S}(x)$ can be decomposed as $d_{S}(x)=\sum_{i=1}%
^{n}d_{S}^{(i)}(x)$ with%
\begin{equation}
d_{S}^{(i)}(x)=\frac{2x}{\pi}\lim_{\varepsilon\rightarrow0}\frac{2g_{i}%
^{2}\operatorname{Im}\left[  \Sigma(p^{2},m_{i}^{2})\right]  +\varepsilon
}{\left(  x^{2}-M_{0}^{2}+\operatorname{Re}\Pi(x^{2})\right)  ^{2}+\left(
\operatorname{Im}\Pi(x^{2})+\varepsilon\right)  ^{2}}\text{ , }\Pi
(x)=\sum_{i=1}^{n}(\sqrt{2}g_{i})^{2}\Sigma(x^{2},m_{i}^{2})\text{ .}%
\end{equation}
We define $a_{i}(t)$ as the Fourier transform of $d_{S}^{(i)}(x)$:
\begin{equation}
a_{i}(t)=\int_{-\infty}^{\infty}\mathrm{dx}d_{S}^{(i)}(x)e^{-ixt}\text{ }%
\end{equation}
Out of $a_{i}(t)$ we define%
\begin{equation}
A_{i}(t)=\left\vert a_{i}(t)\right\vert ^{2}\text{ , }A_{mix,i}(t)=\sum
_{j=1,j\neq i}^{n}\frac{a_{i}(t)a_{j}^{\ast}(t)+a_{i}^{\ast}(t)a_{j}(t)}%
{2}\text{ .%
\'{}%
}%
\end{equation}
The probability that the state $\left\vert S\right\rangle $ decays in the time
interval $(t,t+dt)$ into the $i$-th channel reads $h_{i}(t)dt$, where%
\begin{equation}
h_{i}(t)=-A_{i}^{\prime}(t)-A_{mix,i}^{\prime}(t)\text{ .}%
\end{equation}
It is possible to define the $n(n-1)/2$ ratio functions
\begin{equation}
R_{ij}(t)=\frac{h_{i}(t)}{h_{j}(t)}\text{ with }i<j,\text{ }i=j=1,...n.
\end{equation}
In the BW limit one has $R_{ij}(t)\rightarrow\Gamma_{i}/\Gamma_{j},$ where
$\Gamma_{i}$ is the tree-level decay width in the $i$-th channel. In general,
we expect large deviations from the BW limit. Indeed, many unstable particles
of the Standard Model decay in more than two channels: the analysis of the
temporal behavior can be performed through the formulae derived here. Explicit
calculations represent an outlook for the future.

\end{document}